\documentclass[letterpaper]{article} 
\usepackage[draft]{aaai25}  
\usepackage{times}  
\usepackage{helvet}  
\usepackage{courier}  
\usepackage[hyphens]{url}  
\usepackage{graphicx} 
\usepackage{natbib}  
\usepackage{caption} 
\usepackage{algorithm}
\usepackage{algorithmic}

\usepackage{newfloat}
\usepackage{listings}
\DeclareCaptionStyle{ruled}{labelfont=normalfont,labelsep=colon,strut=off} 
\floatstyle{ruled}
\newfloat{listing}{tb}{lst}{}
\floatname{listing}{Listing}
\usepackage{booktabs}
\usepackage{amsmath}
\usepackage{amssymb}
\usepackage{dsfont}
\usepackage{upgreek}
\allowdisplaybreaks

\title{Empirical Equilibria in Agent-based Economic systems with Learning agents}
\author {
    Kshama Dwarakanath\textsuperscript{\rm 1},
    Svitlana Vyetrenko\textsuperscript{\rm 1},
    Tucker Balch\textsuperscript{\rm 2}
}
\affiliations {
    \textsuperscript{\rm 1}JP Morgan AI Research, San Francisco, California, USA\\
    \textsuperscript{\rm 2}Emory University, Atlanta, Georgia, USA\\
    kshama.dwarakanath@jpmorgan.com
}

\usepackage{bibentry}

\begin{document}

\maketitle

\begin{abstract}
We present an agent-based simulator for economic systems with heterogeneous households, firms, central bank, and government agents. These agents interact to define production, consumption, and monetary flow. Each agent type has distinct objectives, such as households seeking utility from consumption and the central bank targeting inflation and production. We define this multi-agent economic system using an OpenAI Gym-style environment, enabling agents to optimize their objectives through reinforcement learning. 
Standard multi-agent reinforcement learning (MARL) schemes, like independent learning, enable agents to learn concurrently but do not address whether the resulting strategies are at equilibrium. 
This study integrates the Policy Space Response Oracle (PSRO) algorithm, which has shown superior performance over independent MARL in games with homogeneous agents, with economic agent-based modeling. We use PSRO to develop agent policies approximating Nash equilibria of the empirical economic game, thereby linking to economic equilibria. Our results demonstrate that PSRO strategies achieve lower regret values than independent MARL strategies in our economic system with four agent types. This work aims to bridge artificial intelligence, economics, and empirical game theory towards future research.
\end{abstract}

%

\section{Introduction}
Agent-based modeling (ABM) offers significant potential for advancing economics by defining agents and their interactions to produce complex emergent behaviors even from simple rules \cite{macal2005tutorial}. ABMs have been applied in robotics \cite{vorotnikov2018multi}, financial markets \cite{byrd2019abides}, traffic management \cite{adler2005multi}, social networks \cite{gatti2014large}, and recently, in simulating social interactions with LLMs \cite{park2023generative}. 
Prominent economists highlight the benefits of ABMs in modeling complex scenarios accounting for human adaptation and learning \cite{farmer2009economy}. 
The field of Agent-based Computational Economics advocates for ABMs' ability to simulate `turbulent' social conditions unseen in historical data, and to model dynamics out of equilibrium \cite{srbljinovic2003introduction,tesfatsion2006handbook}. 
\cite{hamill2015agent} promotes ABMs for bridging the gap between microeconomics (individual agent modeling) and macroeconomics (aggregate observations at system level).
\cite{arthur2021foundations} highlights the advantages of heterogeneous ABMs as a bottom-up approach towards modeling nuances of the real world more accurately. 

Reinforcement learning (RL) involves an agent learning to act in an uncertain, dynamic environment via trial-and-error to maximize objectives \cite{kaelbling1996reinforcement}. 
In multi-agent settings, learning agents introduce non-stationarity and potential partial observability for each other \cite{busoniu2008comprehensive}. Multi-agent reinforcement learning (MARL) studies such problems by modeling the system as a stochastic game where the environment state evolves based on joint actions of all agents \cite{littman1994markov,hu1998multiagent}. 
Standard MARL techniques have independently learning agents that update their policies simultaneously, disregarding other learning agents. MARL is related to game theory, which studies multiple agents in static or repeated tasks \cite{fudenberg1998theory,bowling2000analysis,hu2003nash}. Prior research has compared independent and coordinated (or joint action) learning in simple, cooperative games \cite{tan1993multi,claus1998dynamics}.

Recent work has incorporated ABMs and MARL in economic systems \cite{trott2021building,curry2022analyzing,mi2023taxai}, but there is little focus on whether learned agents are at equilibrium, a central notion in economics \cite{hahn1973notion}. 
Empirical game-theoretic analysis (EGTA) studies equilibria in empirical games defined by agent interactions in complex ABMs \cite{wellman2020economic}. 
Here, we use RL for strategy generation alongside EGTA to learn approximate Nash equilibrium strategies for our economic ABM. Specifically, we employ the Policy Space Response Oracle (PSRO) algorithm proposed in \cite{lanctot2017unified} to integrate MARL with equilibrium analysis for multiple heterogeneous economic agents. We summarize our contributions below. 
\begin{enumerate}
\item We consider a Python-based macroeconomic ABM with heterogeneous households, firms, a central bank, and government, that allows easy customization of agent configurations and incorporating reinforcement learning capabilities through OpenAI Gym-style environments.
    \item We base agent heterogeneity parameters and action spaces on economic literature and real-world data, such as US wage statistics, to ensure realistic simulation in the absence of publicly available, individual agent data.
    \item We use the PSRO algorithm to learn adaptive strategies for our economic agents, that form part of a Nash equilibrium of this general-sum game involving four agent types, with multiple agents per type.
    \item We compare strategies learned with independent MARL and PSRO in a hypothetical economic scenario investigating the impact of heterogeneous household skills on their preferences to work at different firms. Findings show that PSRO households work more hours at firms where they have higher skills. Aside from qualitative differences in strategies from both schemes, PSRO strategies achieve lower total regret when agents are allowed to unilaterally switch to the alternative strategy.
\end{enumerate}

\section{Literature Review}
\paragraph{Economic Models.}
Dynamic Stochastic General Equilibrium (DSGE) models are macroeconomic models widely used by Central Banks for macroeconomic forecasting and policy analysis \cite{del2013dsge}.
Early work, like \cite{kydland1982time}, introduced a DSGE model with a representative household and firm, analyzing steady-state observables and quadratic approximations around them. \cite{krusell1998income} extended this by incorporating household heterogeneity in income, wealth, and temporal preferences.
Modern macroeconomic modeling focuses on models estimated from real data \cite{christiano2005nominal,woodford2009convergence}. 
The Federal Reserve Bank of New York provides public access to its DSGE model \cite{negro2015}, and forecasts \cite{frbny_dsge_jl}.
Despite extensive literature, DSGE models rely on linearization techniques studying local perturbations around deterministic steady states, along with restrictive representative agent assumptions. These simplifying assumptions limit their ability to capture the full complexity of real economies \cite{haldane2019drawing}, and make them susceptible to model mis-specification errors \cite{farmer2009economy}. 

\paragraph{Agent-based modeling in Economics.}
The field of Agent-based Computational Economics uses ABMs to simulate interactions among economic agents, addressing factors like agent heterogeneity, adaptation and modeling dynamics out of equilibrium unlike DSGE models \cite{tesfatsion2006handbook}. 
ABMs offer a flexible framework to model complex, heterogeneous, bounded rational economic agents with diverse objectives \cite{stiglitz2018modern}.
\cite{deissenberg2008eurace} aimed to create a comprehensive ABM of the European economy, initially focusing on labor market dynamics and later expanding to include industry evolution, credit markets, and consumption \cite{dawid2016heterogeneous}. While calibrating ABMs with real data is challenging due to the numerous agent-specific parameters, they can be validated by reproducing economic stylized facts. 
There is significant work on ABMs of endogenous growth and business cycles which are empirically validated by replicating a set of micro- and macro-economic facts \cite{dosi2006evolutionary,dosi2017micro}.
They use rule-based agents with predefined behaviors, offering a framework for studying monetary and fiscal policy \cite{dosi2015fiscal}.

\paragraph{Agent-based economic modeling and RL.}
RL techniques are well suited to model household behavior in economics, where households seek to maximize utilities over time \cite{chen2021deep,hill2021solving,atashbar2023ai}. 
RL has also been used to optimize other economic strategies, such as revenue redistribution \cite{koster2022human}, monetary policy \cite{chen2021deep,hinterlang2021optimal}, and tax policy \cite{zheng2022}.
\cite{zheng2022} pioneered multi-agent RL (MARL) in economic ABMs by studying tax policy design with agents and a planner. The planner sets marginal tax rates to balance equality and productivity, while agents maximize their endowment utility. 
\cite{curry2022analyzing} introduced a macroeconomic real-business-cycle ABM using MARL for consumers, firms, government strategies, albeit with limitations including absence of the central bank's role in monetary policy, uniform tax redistribution, and omission of inventory holding risk for firms.
\cite{mi2023taxai} developed an economic simulator for taxation using MARL, scaling to 10,000 household agents who maximize consumption utilities, while the government aims to improve social welfare and economic growth. 
Although only a few recent studies explore MARL within economic ABMs, this research area is gradually expanding \cite{dwarakanath2024abides,brusatin2024simulating}. But, none of these studies examine the impact of the MARL scheme on achieving a Nash equilibrium. 
\paragraph{Empirical Game-Theoretic Analysis.}
Game theory studies the interaction between multiple players or agents, each aiming to maximize their utility based on the joint strategy of all agents \cite{von1928theory}.
A mixed strategy Nash equilibrium is a set of probability distributions 
where no agent can gain utility by deviating from their probabilistic strategy, given other agents' strategies \cite{von2007theory}. This concept is crucial for equilibrium in economic systems, as it implies that no agent has an incentive to change their strategy unilaterally.
While there are tools to find Nash equilibria \cite{nashpy,gambit}, applying these to our economic ABM is hard because they need a finite game definition. First, each agent's strategy set is unbounded as strategies are temporal policies, mapping states to actions. Second, analytically characterizing agent utilities is hard as they represent cumulative rewards over time, which depends on other agents' strategies. Empirical game-theoretic analysis (EGTA) offers a solution by addressing games where producing an explicit game model is impractical \cite{walsh2002analyzing,wellman2006methods}. EGTA bridges agent-based simulation and game theory by estimating a game from simulation data generated over a space of strategies \cite{wellman2020economic}.

\paragraph{EGTA and RL.}
In EGTA, building strategy sets for simulation can be done based on heuristic strategies \cite{walsh2002analyzing} or generating new strategies through RL \cite{schvartzman2009stronger}. 
Subsequently, \cite{lanctot2017unified} introduced the Policy Space Response Oracle (PSRO) algorithm, for strategy/\textit{policy space} exploration in general multi-player games using deep RL (the \textit{response oracle}). PSRO iteratively adds new strategies by computing an approximate best response using RL to a meta-strategy (set of mixed strategies across agents) of the empirical game.
Unlike independent MARL, which deals with non-stationarity due to agents updating policies simultaneously, PSRO updates one agent at a time in a stationary environment where others follow their meta-strategy. 
PSRO generalizes the Double Oracle algorithm for 2-player games, when using Nash equilibrium as the meta-strategy \cite{mcmahan2003planning,iter_deep_rl_mason}.
\cite{smith2020iterative} enhanced PSRO using Q-Mixing to ease learning, showing reduced training time for 2 and 3 player games with homogeneous players. 

While previous work has compared independent learning to schemes incorporating joint action information for cooperative games \cite{claus1998dynamics}, little is known about how these techniques compare for dynamic general-sum games on the equilibria reached (if and when they do). In this work, we use deep RL to generate new strategies within the EGTA framework for an economic ABM with households, firms, central bank, and government. Our approach applies PSRO to economic ABMs, setting the meta-strategy to a Nash equilibrium of the empirical game in each epoch. By solving for a Nash equilibrium in every epoch and using RL to improve existing strategies to reach a new Nash equilibrium, we establish a connection to (empirical) economic equilibrium, which is otherwise difficult to elucidate. We also contrast independent MARL (IMARL) and PSRO on the learned strategies, focusing on their regret values.


\section{Multi-Agent Economic System}\label{sec:maes}
Our economic ABM has four types of agents: \textbf{Households} who are consumers of goods and provide skilled labor for production of goods; \textbf{Firms} who utilize labor to produce goods and pay wages; \textbf{Central Bank} that monitors price inflation and production to set interest rate for household savings; and the \textbf{Government} that collects income taxes from households that could potentially be redistributed as tax credits.
Each agent is modeled as a learner aiming to maximize the discounted sum of its rewards over a horizon $H$. This multi-agent economic system is formalized as a Partially Observable Markov Game \cite{hu1998multiagent}, represented by $\Gamma=\langle \mathcal{N},\mathcal{S},\lbrace\mathcal{A}_i\rbrace_{i=1}^n,\lbrace\mathcal{O}_i\rbrace_{i=1}^n,\mathds{T},\lbrace \mathds{O}_i\rbrace_{i=1}^n,\lbrace R_i\rbrace_{i=1}^n,\lbrace\beta_i\rbrace_{i=1}^n$, $H\rangle$ where 
$\mathcal{N}=\lbrace1,2,\cdots,n\rbrace$ is the set of agents, $\mathcal{S}$ is the state space, $\mathcal{A}_i$ is the action space of agent $i$ with $\mathcal{A}=\mathcal{A}_1\times\mathcal{A}_2\times\cdots\times\mathcal{A}_n$ denoting the joint action space, $\mathcal{O}_i$ is the observation space of agent $i$, $\mathds{T}:\mathcal{S}\times\mathcal{A}\rightarrow\mathds{P}\left(\mathcal{S}\right)$ is the transition function mapping current state and joint action to a probability distribution over the next state, $\mathds{O}_i:\mathcal{S}\rightarrow\mathds{P}\left(\mathcal{O}_i\right)$ is the observation function mapping current state to a probability distribution over observations of agent $i$, $R_i:\mathcal{S}\times\mathcal{A}\rightarrow\mathds{R}$ is the reward function of agent $i$, $\beta_i\in[0,1)$ is the discount factor of agent $i$, and $H$ is the horizon.
The objective of each agent $i\in\mathcal{N}$ is to find a sequence of individual actions to maximize expected sum of discounted rewards over the horizon
\begin{align}
    \max_{\left(a_i(0),\cdots,a_i(H-1)\right)}\ \mathds{E}\left[\sum_{t=0}^{H-1}\beta_i^tR_i\left(s(t),a_1(t),\cdots,a_n(t)\right)\right]\nonumber
\end{align}
where $s(t+1)\sim\mathds{T}\left(s(t),a_1(t),\cdots,a_n(t)\right)$. 
Time step $t$ is a simulation period, typically one quarter of a year in economic ABMs. 

\subsection{Households}

Households are the consumer-workers that provide skilled labor for production at firms, while also consuming some of the produced goods. They are paid wages for their labor at the firms and pay for the price of consumed goods. The government collects income taxes on their labor income, part of which could be redistributed back to households as tax credits in the subsequent year. They also earn (accrue) interest on their savings (debt) from the central bank. These monetary inflows and outflows govern the dynamics of household savings from one time step to the next. 

The \textbf{observations} of household $i$ at time $t$ include tax credit $\kappa_{t,i}$, tax rate $\tau_t$, interest rate $r_t$, wages of all firms $\lbrace w_{t,j}:\forall j\rbrace$, prices of goods of all firms $\lbrace p_{t,j}:\forall j\rbrace$ and their monetary savings $m_{t,i}$.
The \textbf{actions} include their hours of labor for all firms $\lbrace n_{t,ij}:\forall j\rbrace$ and units of good requested for consumption at all firms $\lbrace c^{req}_{t,ij}:\forall j\rbrace$. 
The \textbf{dynamics} related to household $i$ are given by\begin{align}
c_{t,ij}&=\min\left\lbrace c_{t,ij}^{req},Y_{t,j}\cdot\frac{c_{t,ij}^{req}}{\sum_kc_{t,kj}^{req}}\right\rbrace\label{eq:H_dyn1}\\
    m_{t+1,i}&=(1+r_t)m_{t,i}+\sum_j\left(n_{t,ij}\omega_{ij}w_{t,j}-c_{t,ij}p_{t,j}\right)\nonumber\\
    &-\tau_t\cdot\sum_jn_{t,ij}\omega_{ij}w_{t,j}+\kappa_{t,i}\label{eq:H_dyn2}
\end{align}
In (\ref{eq:H_dyn1}), when the requested consumption for firm $j$ exceeds its inventory $Y_{t,j}$, goods are allocated proportionally to requests so that the realized consumption of goods of firm $j$ by household $i$ is $c_{t,ij}$.
(\ref{eq:H_dyn2}) is the evolution of savings from $t$ to $t+1$ where $\omega_{ij}$ denotes the skill of household $i$ at firm $j$. 

The \textbf{reward} for household $i$ at $t$ is given by $\sum_{j}u(c_{t,ij},n_{t,ij},m_{t+1,i};\gamma_i,\nu_i,\mu_i)$ where \begin{align}
    u(c,n,m;\gamma,\nu,\mu)&=\frac{c^{1-\gamma}}{1-\gamma}-\nu n^2+\mu\cdot\mathrm{sign}(m)\frac{|m|^{1-\gamma}}{1-\gamma}\nonumber
\end{align}
with an isoelastic utility from consumption and savings, and a quadratic disutility of labor
\cite{evans2005policy}. Households are heterogeneous in their skills per firm $\omega_{ij}$, and utility parameters $(\gamma_i,\nu_i,\mu_i)$.


\subsection{Firms}
Firms are the producer-employers that use household labor to produce goods for consumption. They pay wages for the received labor and receive revenue from prices paid for consumed goods. Their production is subject to an exogenous, stochastic production factor that captures any external shocks \cite{hill2021solving}. Firms accumulate inventory when they produce more goods than consumed by households, which they seek to minimize. 

The \textbf{observations} of firm $j$ at time $t$ include total household labor $\sum_in_{t,ij}\omega_{ij}$, total consumption $\sum_ic_{t,ij}$, exogenous shock $\varepsilon_{t,j}$, exogenous production factor $\epsilon_{t-1,j}$, previous wage $w_{t,j}$, previous price $p_{t,j}$ and inventory $Y_{t,j}$.
The \textbf{actions} include wage per unit of labor $w_{t+1,j}$ and price per unit of good $p_{t+1,j}$ that go into effect at the next time step. 
The \textbf{dynamics} of quantities related to firm $j$ are given by\begin{align}
\epsilon_{t,j}&=\left(\epsilon_{t-1,j}\right)^{\rho_j}\exp\left(\varepsilon_{t,j}\right)\label{eq:F_dyn1}\\
y_{t,j}&=\epsilon_{t,j}\left(\sum_in_{t,ij}\omega_{ij}\right)^{\alpha_j}\label{eq:F_dyn2}\\
Y_{t+1,j}&=Y_{t,j}+y_{t,j}-\sum_ic_{t,ij}\label{eq:F_dyn3}
\end{align}
(\ref{eq:F_dyn1}) describes the exogenous production factor $\epsilon_{t,j}$ as following a log-autoregressive process with coefficient $\rho_j\in[0,1]$, $\epsilon_{0,j}=1$, and $\varepsilon_{t,j}\sim\mathcal{N}\left(\bar{\varepsilon}_{j},\sigma^2_{j}\right)$ being an exogenous shock.
(\ref{eq:F_dyn2}) is the firm's production process per a Cobb-Douglas production function using skilled labor with elasticity parameter $\alpha_j\in[0,1]$ \cite{cobb1928theory}. 
The firm updates its inventory at the next time step based on current inventory and the difference between supply and demand as in (\ref{eq:F_dyn3}).

The \textbf{reward} for firm $j$ at $t$ is given by \begin{align}
    p_{t,j}\sum_{i}c_{t,ij}-w_{t,j}\sum_in_{t,ij}\omega_{ij}-\chi_jp_{t,j}Y_{t+1,j}\nonumber
\end{align}
where the first two terms depict profits as the difference in consumption revenue and labor costs, while the last term captures inventory risk. Firms are heterogeneous in their sector, modeled by shock process and production parameters $(\rho_j,\bar{\varepsilon}_{j},\sigma_{j},\alpha_j)$, and their weighting for inventory risk $\chi_j$.


\subsection{Central Bank}

The central bank is the regulatory agency that monitors the prices and production of goods to set interest rates for household savings. By changing the interest rate on household savings, it affects the consumption and labor patterns of the household. These in turn affect the prices of goods produced by firms. The central bank seeks to set interest rates to meet inflation targets and boost production.

The \textbf{observations} of the central bank at time $t$ include total price of goods over the last five quarters $\lbrace \sum_jp_{t-k,j}:\forall k\in\{0,1,2,3,4\}\rbrace$ and total firm production $\sum_jy_{t,j}$.
The \textbf{action} includes the interest rate $r_{t+1}$ that goes into effect at the next time step. 
The \textbf{dynamics} include computing the annual inflation rate of total price as $\pi_t=\frac{\sum_jp_{t,j}}{\sum_jp_{t-4,j}}$.

The \textbf{reward} for the central bank is given by \begin{align}
    -\left(\pi_t-\pi^\star\right)^2+\lambda\left(\sum_jy_{t,j}\right)^2\nonumber
\end{align}
where $\pi^\star$ is the target inflation rate. And, $\lambda>0$ weighs the production reward in relation to meeting the inflation target.

\subsection{Government}
The government is the regulatory agency that collects taxes from households on their labor income in order to maintain infrastructure. It sets an income tax rate and can choose to distribute a portion of the collected taxes back to households as tax credits in order to improve household social welfare.

The \textbf{observations} of the government at time $t$ include the previous tax rate $\tau_t$, previous tax credits $\lbrace\kappa_{t,i}:\forall i\rbrace$, previous tax collected $\lbrace\tau_t\sum_jn_{t,ij}\omega_{ij}w_{t,j}:\forall i\rbrace$ and a time varying household weight in relation to social welfare $\lbrace l_{t,i}:\forall i\rbrace$. 
Our framework lets the designer choose weights $l_{t,i}$ based on their choice of social welfare metric e.g., $l_{t,i}\equiv1$ for the utilitarian social welfare function versus $l_{t,i}=\mathds{1}\lbrace i=\arg\min_km_{t,k}\rbrace$ for the Rawlsian social welfare function. We choose $l_{t,i}$ to be a function of household savings with parameters $\alpha_l>0,\beta_l>0$, and $l_2>l_1>0$ as
$l_{t,i}=\begin{cases}
        \max\lbrace l_1,-\alpha_lm_{t,i}+\beta_l\rbrace,\textnormal{ if }m_{t,i}>0\\
        \min\lbrace l_2,-2\alpha_lm_{t,i}+\beta_l\rbrace,\textnormal{ if }m_{t,i}\leq0\\
    \end{cases}$.
So, the government underweights high-savings households and overweights those with high debt, 
ensuring all households are weighted between $l_1$ and $l_2$.
The \textbf{actions} include tax rate $\tau_{t+1}$, and fraction of tax credit distributed to each household $i$, $f_{t+1,i}$, that go into effect at the next time step. 
The \textbf{dynamics} are given by $\kappa_{t+1,i}=\xi f_{t+1,i}\sum_k \left(\tau_t\sum_jn_{t,kj}\omega_{kj}w_{t,j}\right)$, where $f_{t,i}\in[0,1]$ with $\sum_if_{t,i}=1$ so that portion $\xi\in[0,1]$ of collected taxes is redistributed. The tax credit for household $i$ at $t+1$ is a fraction $f_{t+1,i}$ of this redistributed amount.

The \textbf{reward} for the government is a measure of household social welfare, given here by a weighted sum of their utilities \begin{align}
    \sum_il_{t,i}R_{t,i,\mathbf{H}}\nonumber
\end{align}
where $l_{t,i}$ is the weight associated to household $i$ and, $R_{t,i,\mathbf{H}}=\sum_{j}u(c_{t,ij},n_{t,ij},m_{t+1,i};\gamma_i,\nu_i,\mu_i)$ is the reward function measuring the utility for household $i$ at time $t$. 


\section{Policy Space Response Oracles}
We adapt the Policy Space Response Oracles (PSRO) algorithm from \cite{lanctot2017unified} to our economic ABM in Algorithm \ref{alg:psro}. 
The object of analysis is a Normal form game $\left(\Pi,U,n\right)$ between agents in our $n$-agent economic system. Agent $i$ selects a \textit{pure-strategy}/policy $\uppi_i:\mathcal{O}_i\rightarrow\mathds{P}(\mathcal{A}_i)$ from its policy set $\Pi_i$. The joint policy set is $\Pi=(\Pi_1,\cdots,\Pi_n)$, and the utility function $U:\Pi\rightarrow\mathds{R}^n$ maps joint policy $\uppi=(\uppi_1,\cdots,\uppi_n)$ to utilities (discounted cumulative rewards) for each agent $i$ as $U_i(\uppi)=\mathds{E}\Big[\sum_{t=0}^{H-1}\beta_i^tR_i\left(s(t),a_1(t),\cdots,a_n(t)\right)\Big|$ $s(t+1)\sim\mathds{T}\left(s(t),a_1(t),\cdots,a_n(t)\right),a_j(t)\sim\uppi_j\Big]$.
Agent $i$ can also adopt a \textit{mixed-strategy} $\upsigma_i\in\mathds{P}(\Pi_i)$, randomizing over policies in $\Pi_i$ according the distribution $\upsigma_i$. A joint strategy $\upsigma=(\upsigma_1,\cdots,\upsigma_n)$ is a mixed-strategy Nash equilibrium if no agent $i$ can gain in utility by deviating from $\upsigma_i$ when all other agents play $\upsigma_{-i}$.
Algorithm \ref{alg:psro} sets the meta-strategy to be a Nash equilibrium of the latest empirical game.

\begin{algorithm}[tb]
    \caption{PSRO \protect\cite{lanctot2017unified}
    }
    \label{alg:psro}
    \begin{algorithmic}[1] 
    \REQUIRE Initial policy sets $\Pi_i=\lbrace\uppi_i^0\rbrace$ for every agent $i$.
    \ENSURE PSRO epochs $N$ and deep RL episodes $M$.
        \STATE Simulate utilities $U(\uppi_1^0,\cdots,\uppi_n^0)$
        \STATE Initialize Nash equilibrium $\upsigma_i=\mathbf{1}(\uppi_i^0)$ for every $i$
        \WHILE{Epoch $e$ in $1,\cdots,N$}
        \FOR{Agent $i$ in $1,\cdots,n$\label{alg:agent_type}}
        \FOR{RL episode $k$ in $1,\cdots,M$}
        \STATE Sample other agent policies $\uppi_{-i}\sim\upsigma_{-i}$
        \STATE Train oracle $\uppi_i^e$ over trajectory from $(\uppi_i^e,\uppi_{-i})$
        \ENDFOR
        \STATE Augment policy set $\Pi_i=\Pi_i\cup\lbrace\uppi_i^e\rbrace$
        \ENDFOR
        \STATE Simulate utilities $U(\uppi)$ for new policies $\uppi\in\Pi$
        \STATE Compute Nash equilibrium $\upsigma$ from $U(\Pi)$
        \ENDWHILE
    \end{algorithmic}
    \textbf{Output:} Nash equilibrium $\upsigma=(\upsigma_1,\cdots,\upsigma_n)$.
\end{algorithm}

To compare the quality of strategies got by IMARL and PSRO, we compute regret values against a deviation set containing strategies of both methods. As in \cite{smith2020iterative}, define the total regret of joint strategy $\upsigma=(\upsigma_1,\cdots,\upsigma_n)$ with respect to deviation set $\mathtt{\Sigma}$ as
\begin{align*}
    \mathcal{R}(\upsigma,\mathtt{\Sigma})=\sum_{i=1}^n\left(\max_{\Tilde{\upsigma}_i\in\mathtt{\Sigma}}U_i\left(\Tilde{\upsigma}_i,\upsigma_{-i}\right)-U_i\left(\upsigma_i,\upsigma_{-i}\right)\right)
\end{align*}
where the term within the summation captures regret to agent $i$. Since $\mathtt{\Sigma}=\mathtt{\Sigma}^{\textnormal{IMARL}}\cup\mathtt{\Sigma}^{\textnormal{PSRO}}$, this metric assesses utility loss when agents are confined to one method over the other. As IMARL gives pure strategies, $\upsigma^{\textnormal{IMARL}}=(\mathds{1}\lbrace\uppi_1^{\textnormal{IMARL}}\rbrace,\cdots,\mathds{1}\lbrace\uppi_n^{\textnormal{IMARL}}\rbrace)$.

\section{Experimental Framework}

\subsection{Economic System Simulator}
Our simulator is built on ABIDES, an agent-based interactive discrete event simulator widely used to simulate financial markets with diverse traders \cite{byrd2019abides}. 
ABIDES agent have access to their internal states, and receive information about other agents via messages. A simulation kernel handles message passing, and runs simulations over a specified time horizon while maintaining agent timestamps. 
ABIDES was extended for single agent RL using an OpenAI Gym style extension in \cite{amrouni2021abides}. We expand this to the MARL setting by having a single Gym agent control action setting of all RL agents, allowing for a mix of learning and rule-based agents. 

\paragraph{Agent configuration.}There are four agent classes, one per type as in Section \ref{sec:maes}, each initialized with default heterogeneity parameters sourced from literature. Parameters for household agents follow \cite{chen2021deep}, firm agents follow \cite{hill2021solving}, and central bank follow \cite{hinterlang2021optimal}. See the appendix for more details. Simulation runs require specification of the horizon in quarters, number of agents per type, and heterogeneity parameters if different from the default values.

\paragraph{Agent communication.}Since agents can only access their internal states, any information from other agents must be requested using messages. Recipients of messages respond by sharing a part of their internal states with the sender. 
For example, the household agent sends a message to each firm agent asking for its price and wage. The firm agent responds by sending that information which is used in the household's observation. This applies to every feature in an agent's observation that is external to itself, and pertains to all agents.

\paragraph{Temporal progression.}
At $t=0$, households have \$0 savings, firms have 0 units of inventory and set default prices and wages, the central bank sets default interest rate, and the government sets default tax rate and gives \$0 of tax credits. 
For each subsequent step $t\geq0$, households take in observations to set labor hours and request consumption. Firms use labor to produce goods (\ref{eq:F_dyn1})-(\ref{eq:F_dyn2}), fulfil consumption (\ref{eq:H_dyn1}) and update inventory (\ref{eq:F_dyn3}). They then sets price, wage for the next step. Each household updates savings based on realized consumption (\ref{eq:H_dyn2}) and pays taxes to the government (\ref{eq:H_dyn2}). The central bank monitors prices and productions to set interest rate for the next step. The government collects taxes to set tax rate and distribute credits in the next step.


\paragraph{Calibration and realism.}While the utility of our ABM framework lies in qualitative analysis of economic scenarios rather than forecasting, we take three steps to ensure simulator validity. 
First, we base agent parameters on existing literature. Second, our action spaces reflect typical real-world values from the US Bureau of Labor Statistics \cite{bls} e.g., 
wages chosen from $\lbrace7.25, 19.65, 32.06, 44.46, 56.87\rbrace$, with $\$7.25$ per hour as the minimum wage and $\$32.06$ as the default. 
Lastly, we validate our simulator by reproducing key economic stylized facts, such as the inverse relationship between firm price and consumption and the direct relationship between inflation and interest rates \cite{svensson2020monetary}. Also, \cite{dong2023analyzing} verifies that our simulated data matches real-world trends in the Child Tax Credits scenario.
See the appendix for more details. 

\subsection{Scenario: Heterogeneity in Household Skills}\label{subsec:scenario}
Every household $i$ has a distinct skill level per firm $j$, denoted $\omega_{ij}$. We explore the impact of household skills on their labor preferences at firms.
Consider an economy with 2 heterogeneously skilled households, 2 heterogeneous firms, central bank and government over a horizon of 10 years (40 quarters), where all agents use RL. Firm 1 is a less labor-intensive technology firm, while firm 2 is a more labor-intensive agriculture firm. Household 1 is more skilled at firm 1, with both households having similar skills for firm 2. Households have $\gamma=0.33$, $\nu=0.5$, $\mu=1.0$ and $\beta_{\mathbf{H}}=0.99$ with skills $\begin{bmatrix}
    \omega_{11}&\omega_{12}\\
    \omega_{21}&\omega_{22}
\end{bmatrix}=\begin{bmatrix}
    2&1\\
    1&1
\end{bmatrix}$. Firms have $\beta_{\mathbf{F}}=0.99$, $\rho=0.97$, $\bar\varepsilon=0$, $\sigma=0.1$, and $\chi=0.1$. Firm 1 being less labor intensive has production elasticity $\alpha_1=\frac{2}{3}$, while firm 2 has $\alpha_2=1$. Central bank's parameters are $\pi^\star=1.02$, $\lambda=0.25$ and $\beta_{\mathbf{CB}}=0.99$. 
The government redistributes 10\% of taxes as credits, with $\xi=0.1$, $\beta_{\mathbf{G}}=0.99$, $\alpha_l=1$, $\beta_l=1.2$, $l_1=10^{-3}$ and $l_2=\beta_l+2\alpha_l=3.2$.
\begin{figure*}[t]
    \centering
    \includegraphics[width=0.38\linewidth]{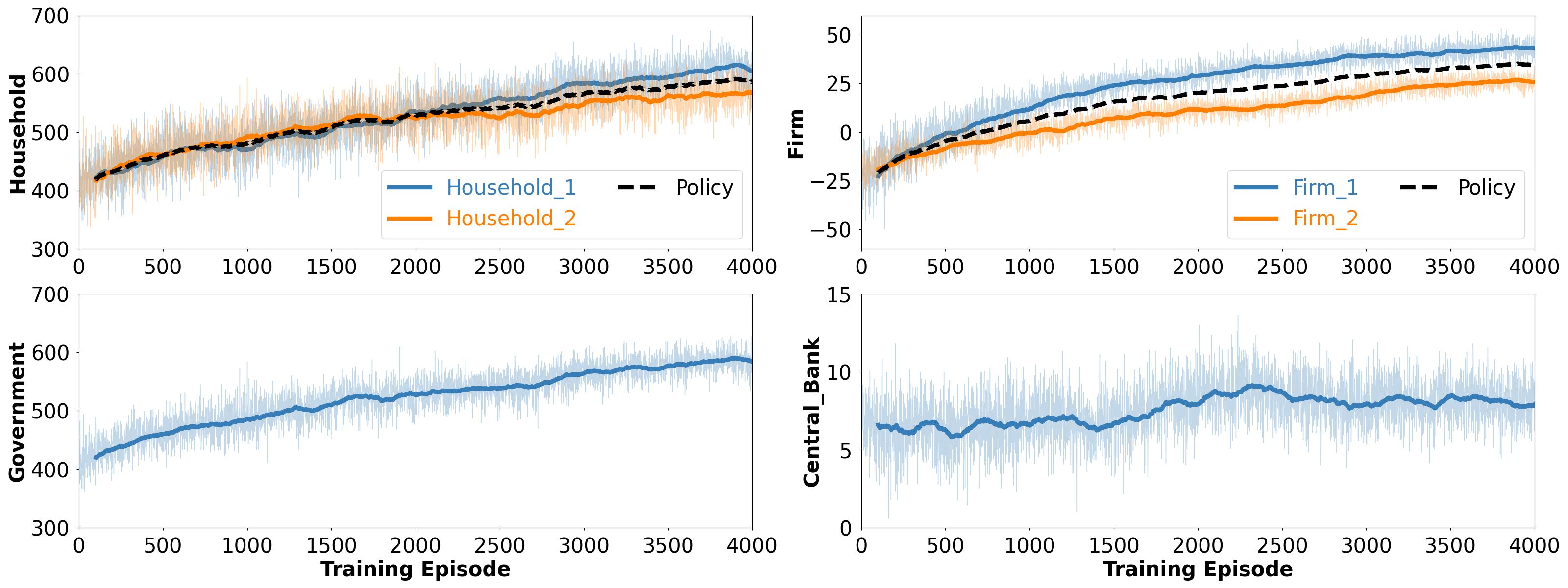}\hspace{0.05\linewidth}
    \includegraphics[width=0.38\linewidth]{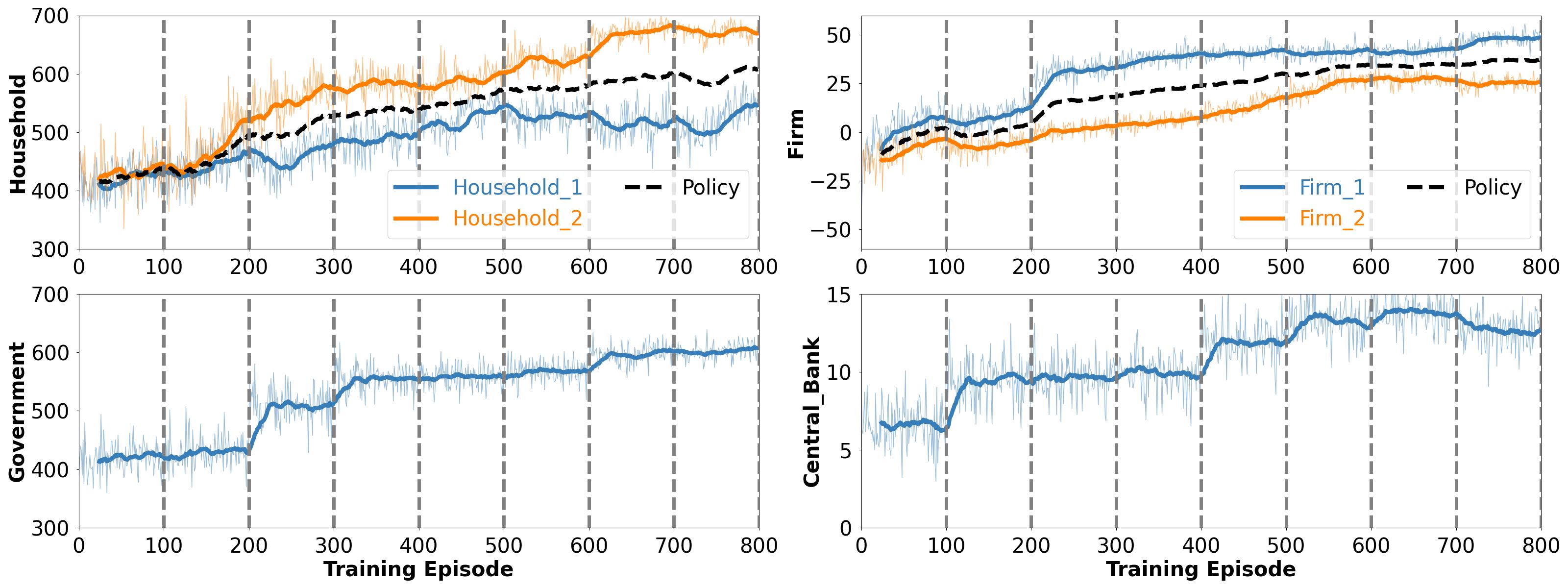}
    \caption{Discounted cumulative rewards during training for IMARL (left) and PSRO (right).}
    \label{fig:training_rewards}
\end{figure*}

\subsection{Learning details}
We use a single policy network per agent type that also inputs heterogeneity parameters to distinguish between agents of the same type. 
The Proximal Policy Optimization (PPO) algorithm within the RLlib package \cite{ppo,rllib} is used to learn policies within PSRO. PPO is also used within IMARL as a baseline for comparison.
Both methods use normalized observations and rewards to stabilize learning.  
Learning rates are $2\times10^{-3}$, $2\times10^{-3}$, $2\times10^{-3}$, $5\times10^{-3}$ for the household, firm, central bank and government policies within PSRO, and are $2\times10^{-3}$, $5\times10^{-3}$, $5\times10^{-3}$, $1\times10^{-2}$ for IMARL. The $n=4$ policies are trained over $M=100$ episodes over $N=8$ PSRO epochs. Our choices for $(N,M)$ are restricted by the need to simulate utilities for $O(e^3)$ new joint policies in epoch $e\in\lbrace1,\cdots,N\rbrace$ over $10$ runs. We parallelize runs across agent types per epoch in line \ref{alg:agent_type} of Algorithm \ref{alg:psro}. See the appendix for additional learning details.

\section{Experimental Results}

\subsection{Comparison of training rewards}

Figure \ref{fig:training_rewards} shows discounted cumulative rewards during training for IMARL (left) and PSRO (right) with each subplot corresponding to agents of a particular type. Faint lines show per episode rewards, solid lines show their moving average, and dashed black lines show rewards of the shared policy for that type. Dashed gray vertical lines on the PSRO plot separate one PSRO epoch from the next.
Notably, PSRO achieves comparable rewards to IMARL, reaching similar levels after $800$ episodes as IMARL does after $4000$ episodes. And, the PSRO policy for the central bank outperforms its IMARL counterpart.
However, note that simulating utilities of new joint policies at the end of each PSRO epoch becomes increasingly computationally intensive as epochs progress. 


\subsection{Comparison of learned policies}\label{subsec:comp_policies}
Agent policies learned with IMARL and PSRO are tested in 500 episodes. And, histograms of agent observations and actions per time step across test episodes are plotted in subsequent figures where vertical lines indicate mean values.

\begin{figure}[tb]
    \centering
    \includegraphics[width=0.38\linewidth]{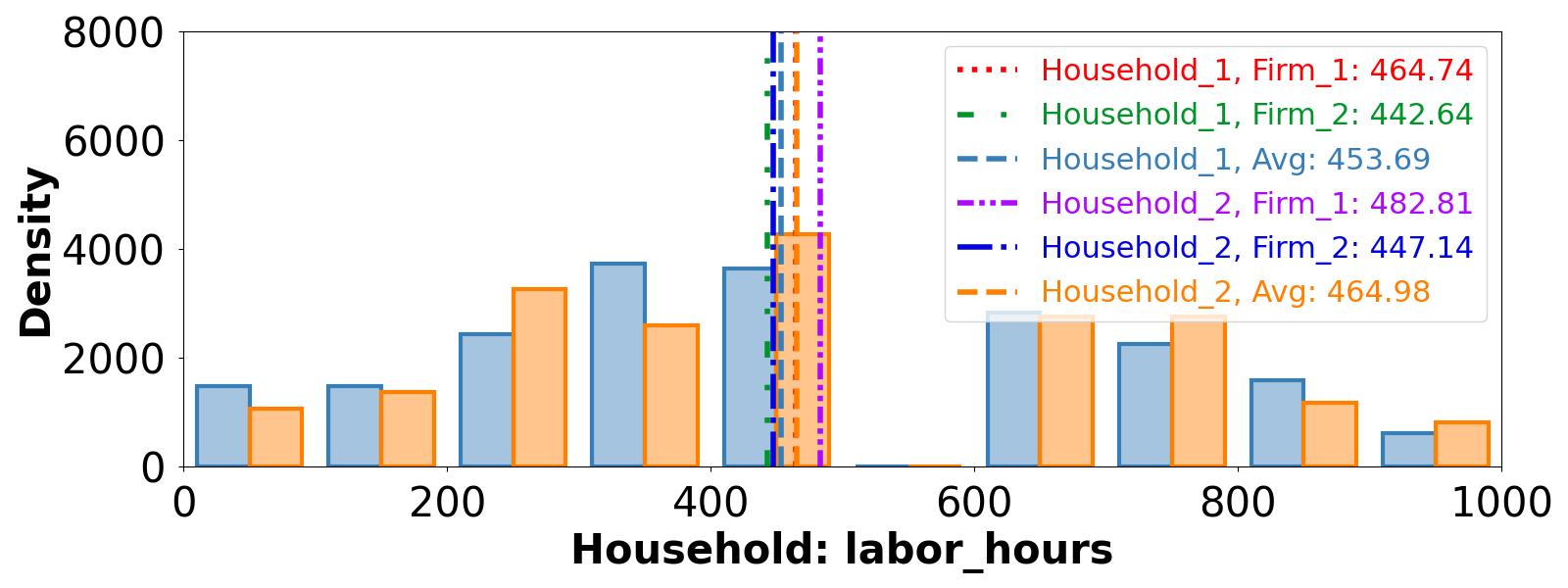}\hspace{0.05\linewidth}
    \includegraphics[width=0.38\linewidth]{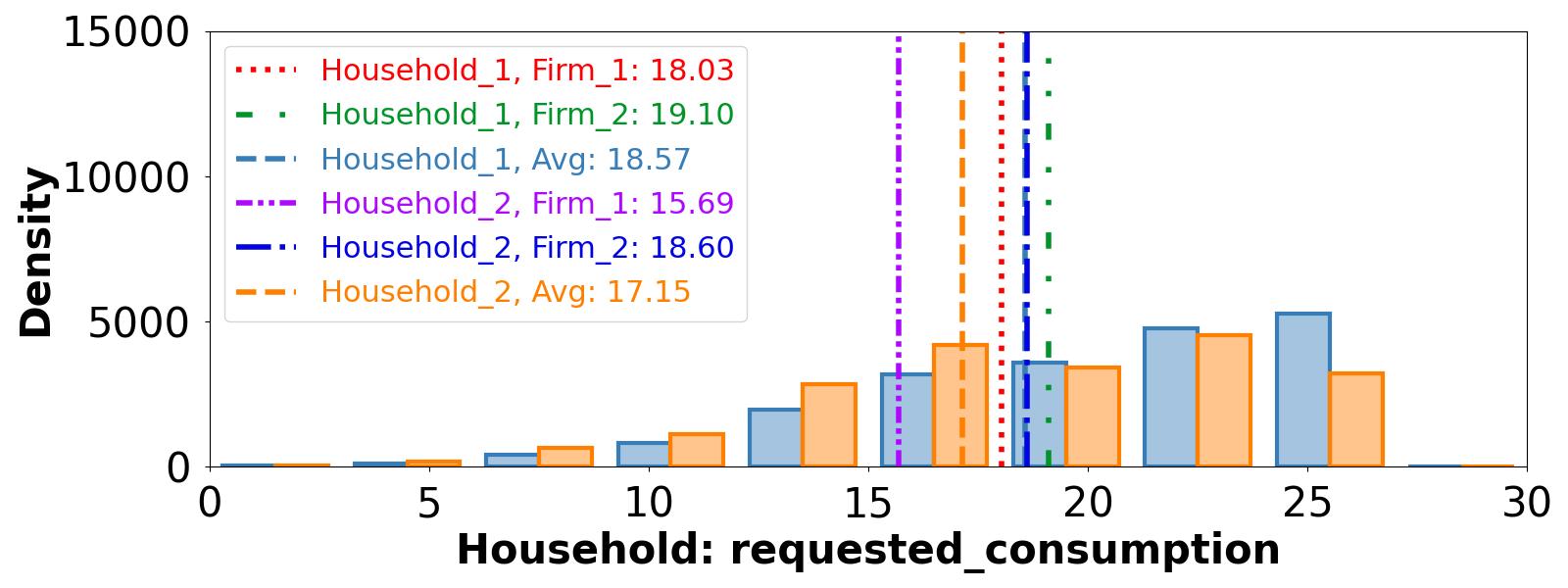}
    \includegraphics[width=0.38\linewidth]{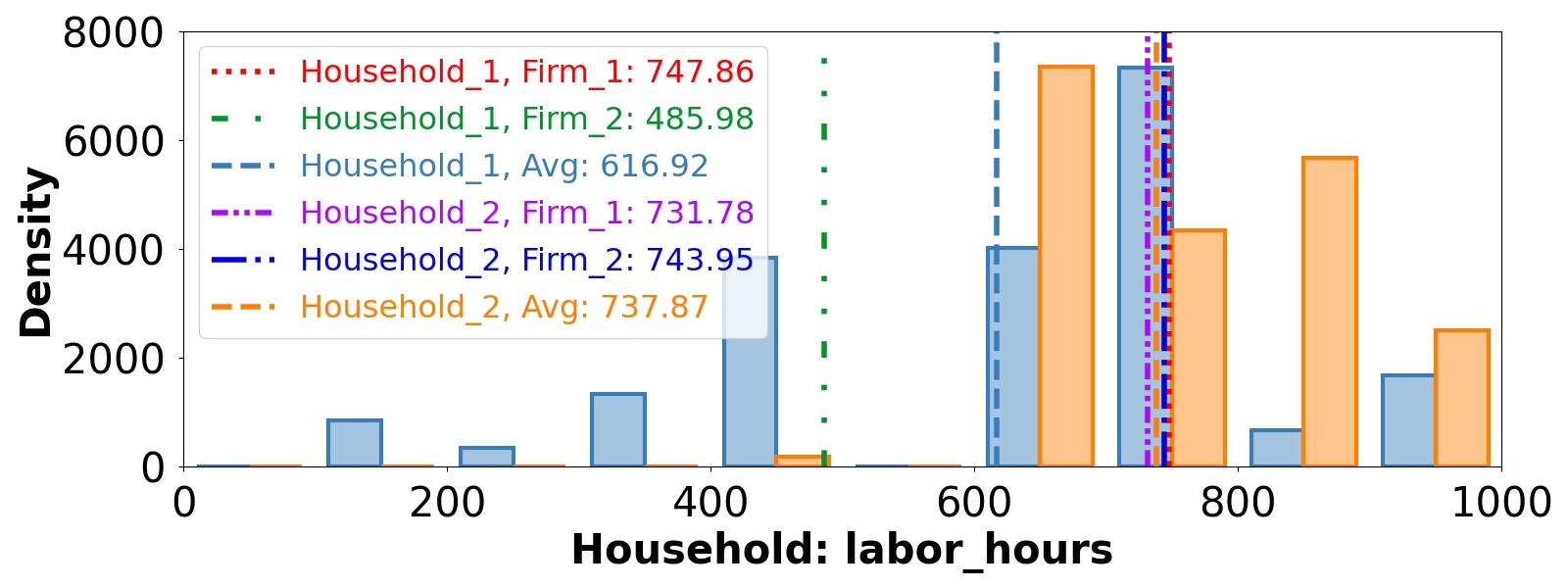}\hspace{0.05\linewidth}
    \includegraphics[width=0.38\linewidth]{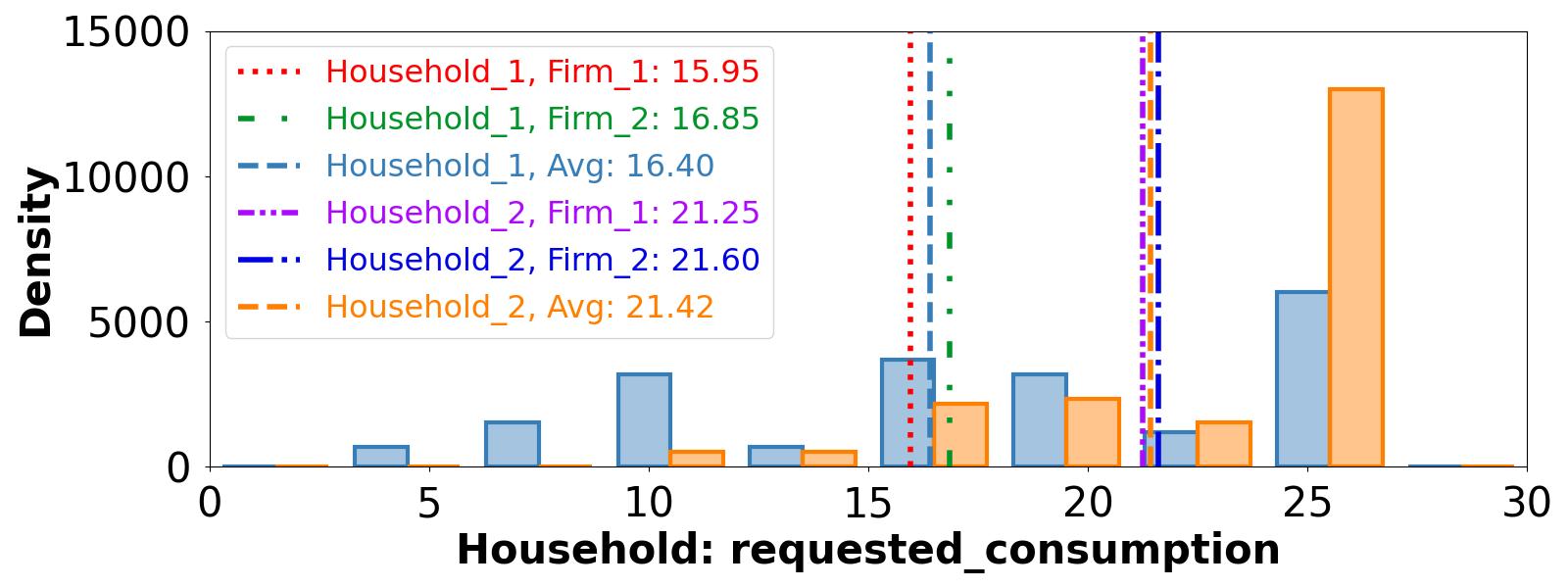}
    \caption{Household labor hours (left) and consumption (right) with IMARL (top) and PSRO (bottom). With PSRO, less-skilled H2 works more and consumes more than H1. 
    }
    \label{fig:household_policy}
\end{figure}
\begin{figure}[tb]
    \centering
    \includegraphics[width=0.38\linewidth]{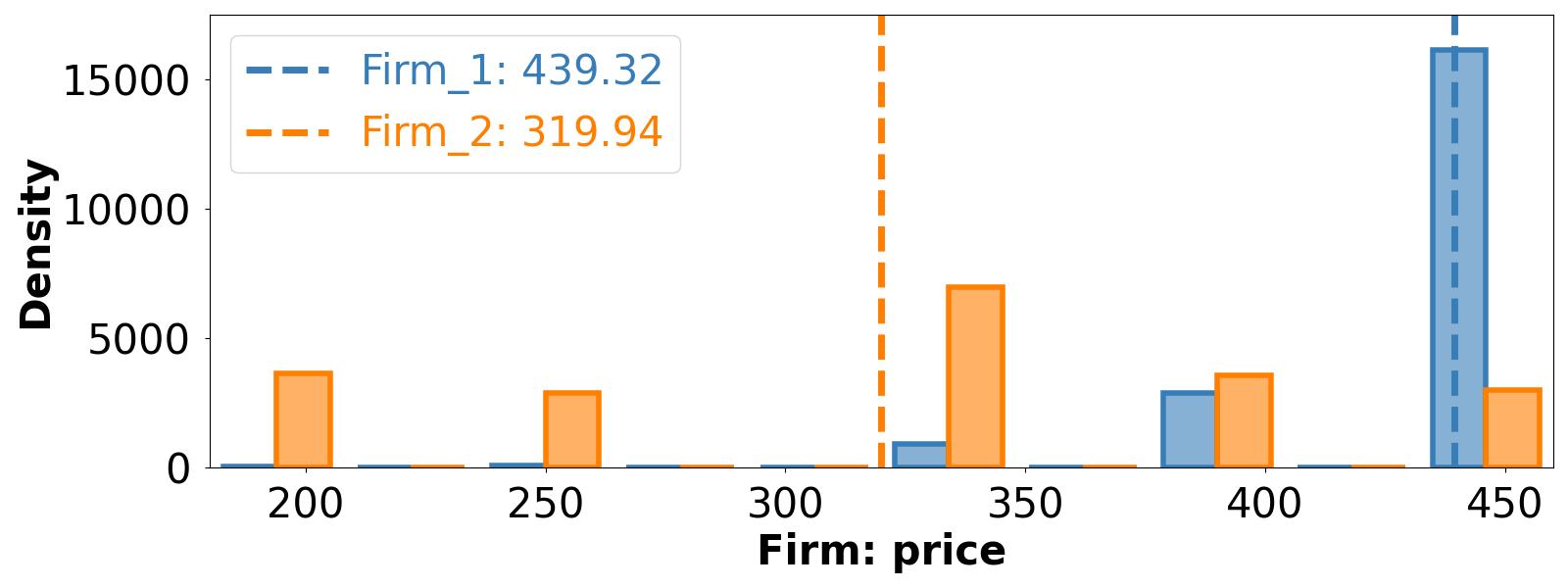}\hspace{0.05\linewidth}
    \includegraphics[width=0.38\linewidth]{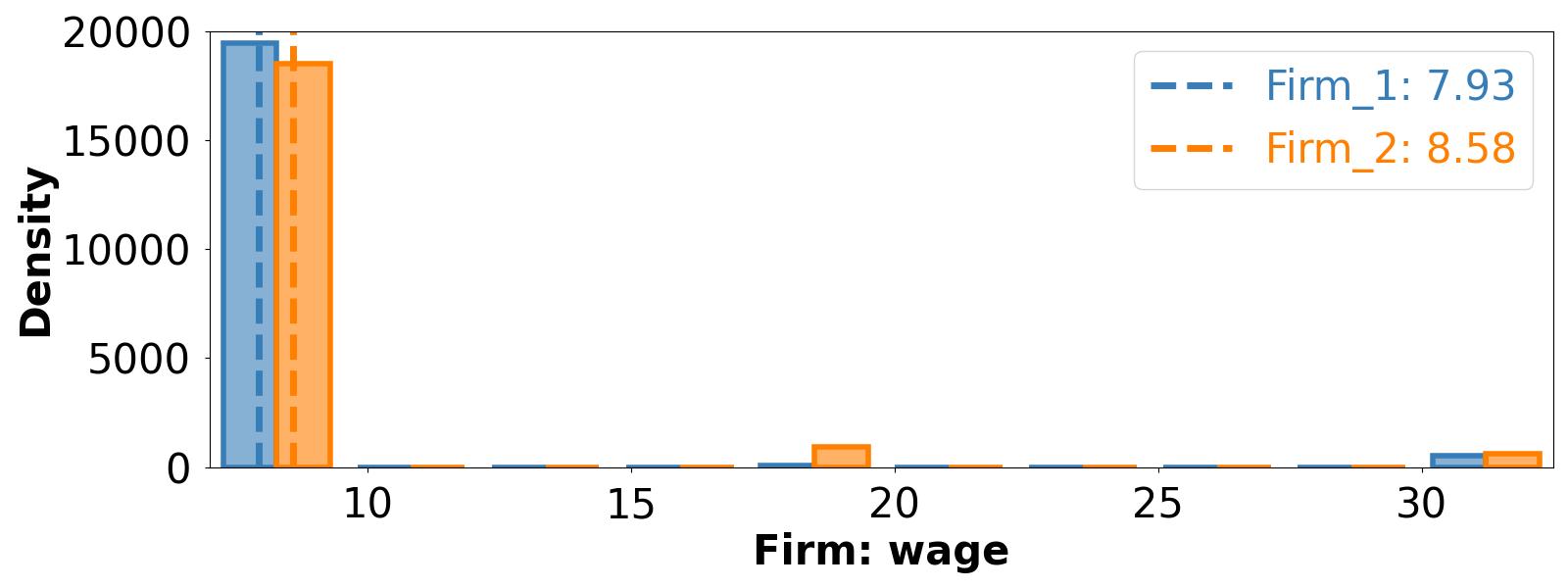}
    \includegraphics[width=0.38\linewidth]{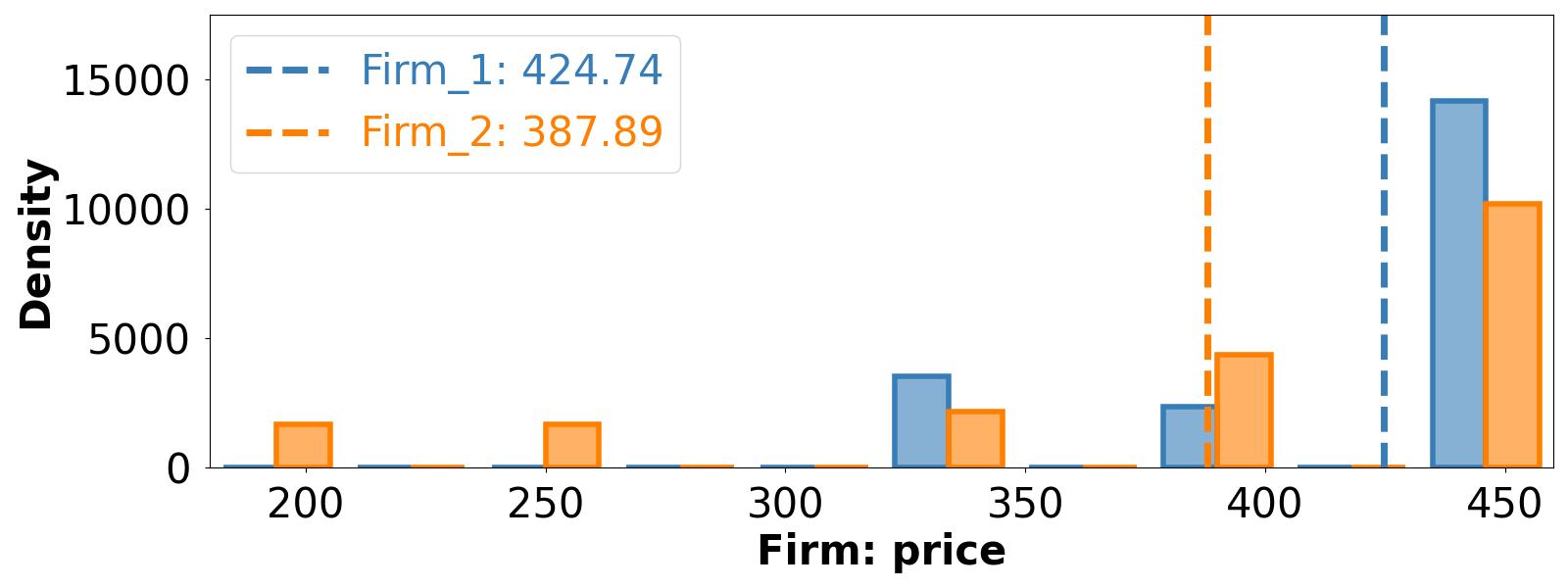}\hspace{0.05\linewidth}
    \includegraphics[width=0.38\linewidth]{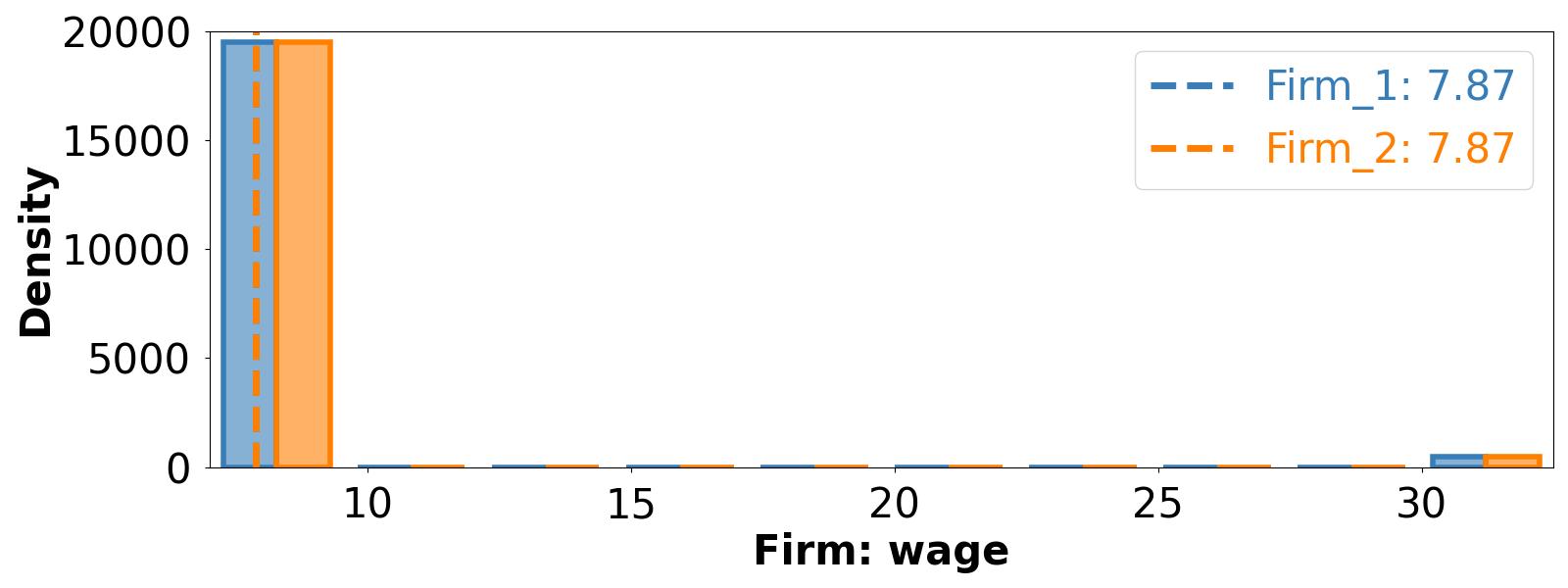}
    \caption{Price (left) and wage (right) of firms with IMARL (top) and PSRO (bottom). While F2 is cheaper than F1 in both schemes, the contrast is lower with PSRO.}
    \label{fig:firm_policy}
\end{figure}
\begin{table*}[t]
    \centering
    \begin{tabular}{cccccc}\toprule
        & Household & Firm & Central Bank & Government & Total\\\midrule
        IMARL & 0.00 (0.00\%) & 3.82 (5.33\%) & 0.00 (0.00\%) & 0.39 (0.06\%) & 4.21 (0.22\%) \\
        PSRO & 0.00 (0.00\%) & 0.00 (0.00\%) & 0.17 (1.27\%) & 0.00 (0.00\%) & 0.17 (0.01\%)\\\bottomrule
    \end{tabular}
    \caption{
    Absolute (and percentage) regret across agent types, with PSRO achieving lower total regret compared to IMARL.}
    \label{tab:regret}
\end{table*}

\paragraph{Household policy.}
Figure \ref{fig:household_policy} compares labor hours (left) and consumption requests (right) of households with IMARL (top) and PSRO (bottom). While both households work similar hours with IMARL, household 1 works fewer hours than household 2 with PSRO. Thus, the impact of lower skill of household 2 forcing it work longer hours to achieve rewards is explicit under PSRO. With PSRO, household 1 being more skilled at firm 1, works more hours there (red line) than at firm 2 (green line), while household 2, equally skilled at both firms, shows no such preference. 
Similarly, while both households request similar consumption with IMARL, household 1 requests less than household 2 with PSRO. Also, both households consume more from firm 2 than from firm 1 under both schemes, a trend influenced by firm pricing as discussed next. The higher contrast in household behaviors with PSRO is also seen in the larger gap between their training rewards in Figure \ref{fig:training_rewards}.

\paragraph{Firm policy.}
Figure \ref{fig:firm_policy} compares price (left) and wage (right) of firms with IMARL (top) and PSRO (bottom).
In both schemes, firm 2 sets lower prices than firm 1 due to its higher production output given the same labor input (see (\ref{eq:F_dyn2}) for elasticity $\alpha_j$), leading to larger inventory accumulation. To reduce inventory, firm 2 prices its goods lower, driving households to prefer consuming from firm 2 over firm 1. The difference in prices is lower under PSRO than IMARL, which can be attributed to the equilibrium behavior of households, who continue consuming more from firm 2 as long as its price remains lower than firm 1's. Thus, firm 2 has no incentive to keep prices very low as long as they're lower than firm 1's, and households have no reason to alter their consumption strategy at equilibrium. About wages, firm 2 offers slightly higher wages than firm 1 under IMARL to encourage more consumption and clear inventory. However, both firms offer identically low wages under PSRO.

\begin{figure}[tb]
    \centering
    \includegraphics[width=0.38\linewidth]{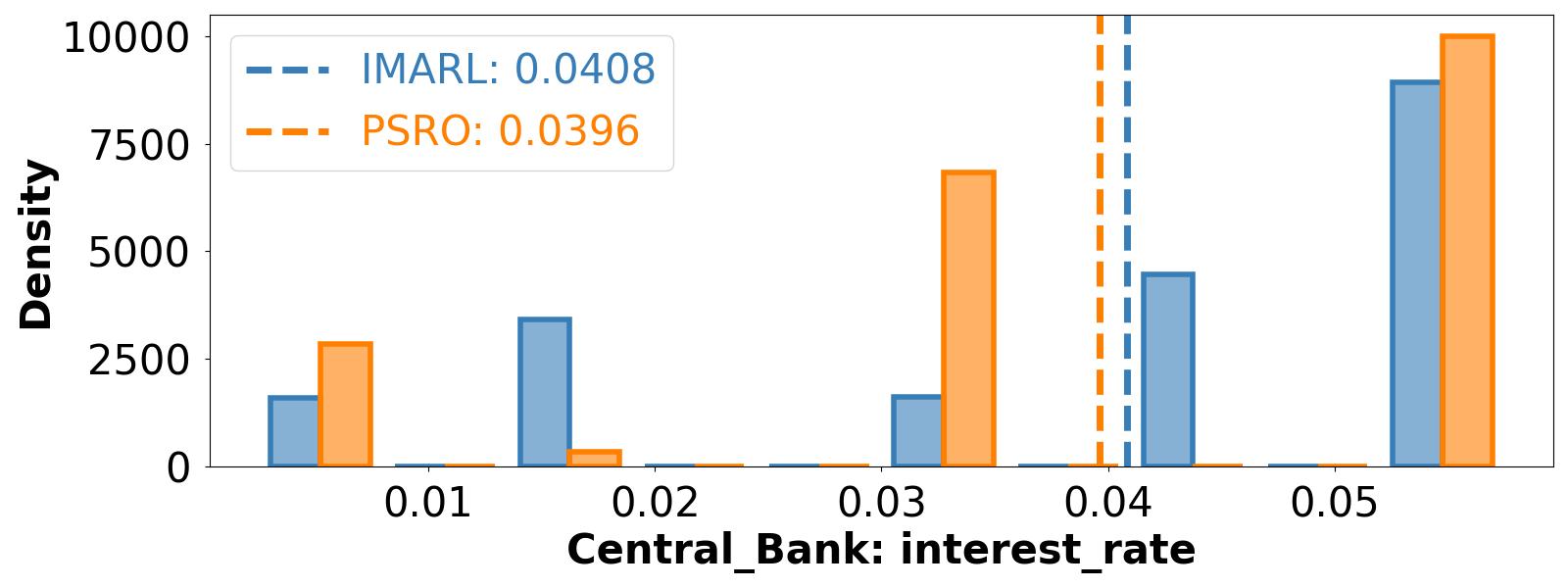}\hspace{0.05\linewidth}
    \includegraphics[width=0.38\linewidth]{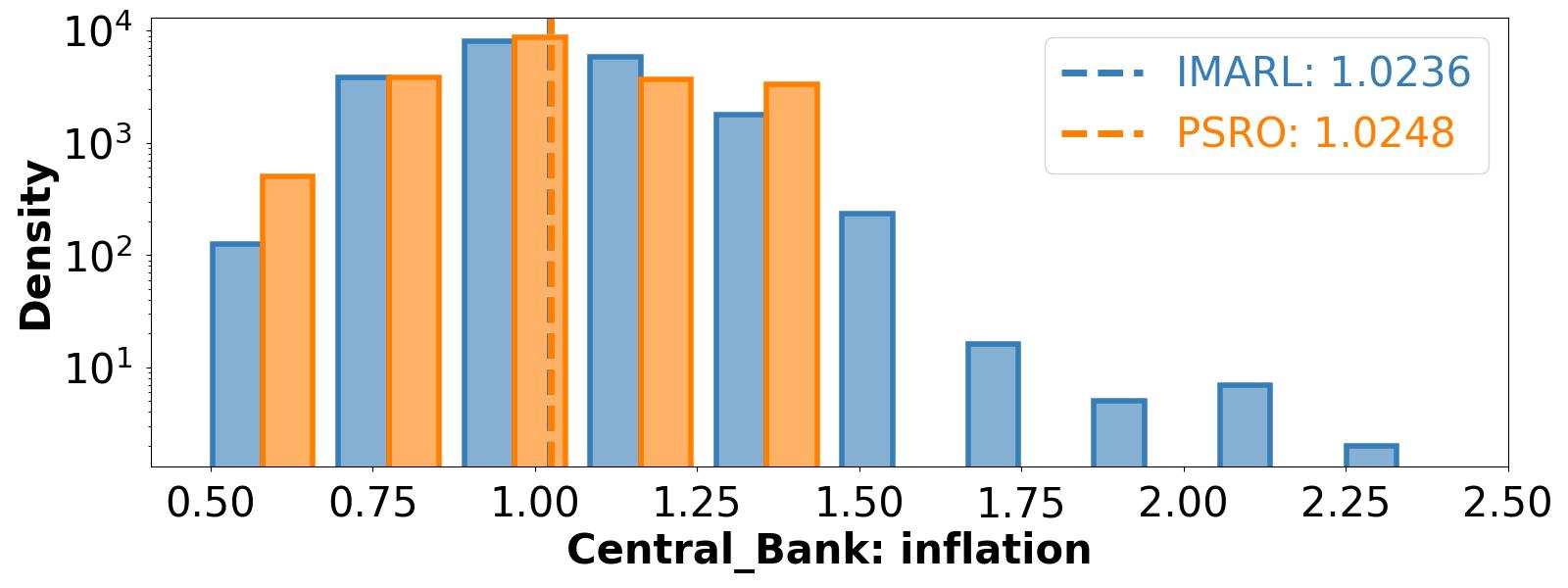}
    \caption{Interest rate set by the central bank (left) and inflation (right) with IMARL and PSRO. 
    PSRO uses fewer rate options than IMARL, while achieving target inflation.}
    \label{fig:cb_policy}
\end{figure}
\begin{figure}[tb]
    \centering
    \includegraphics[width=0.38\linewidth]{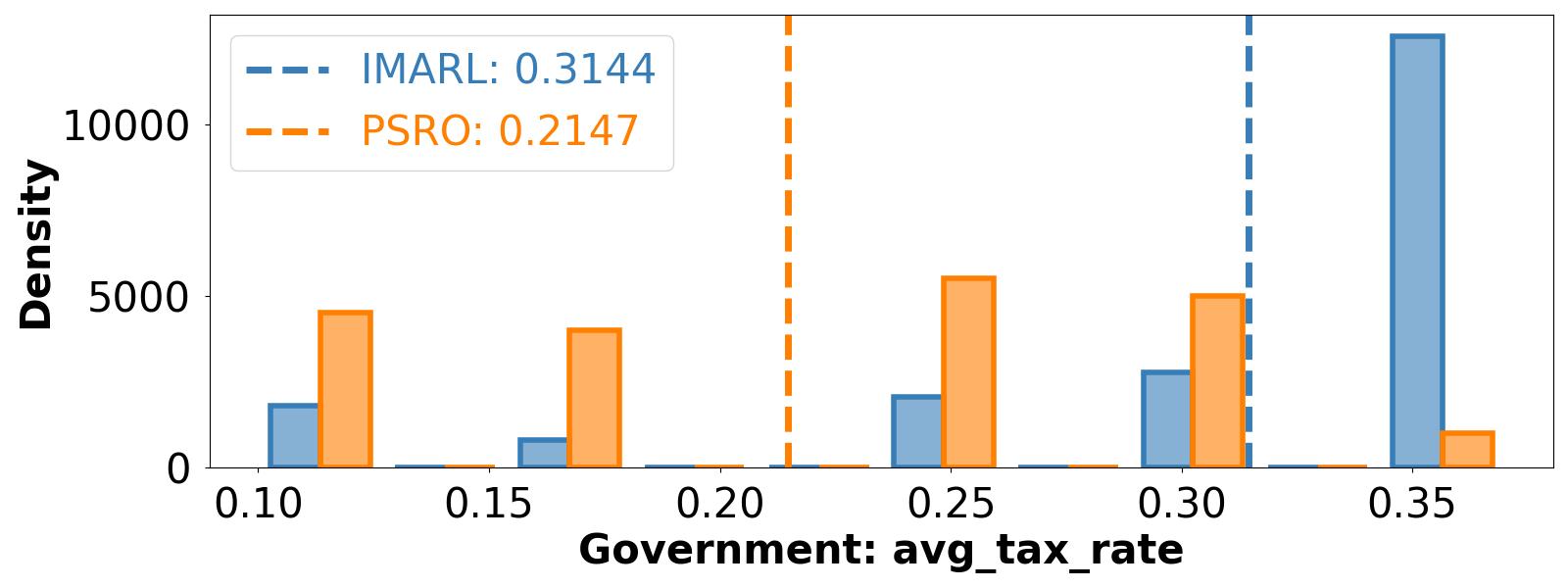}\hspace{0.05\linewidth}
    \includegraphics[width=0.38\linewidth]{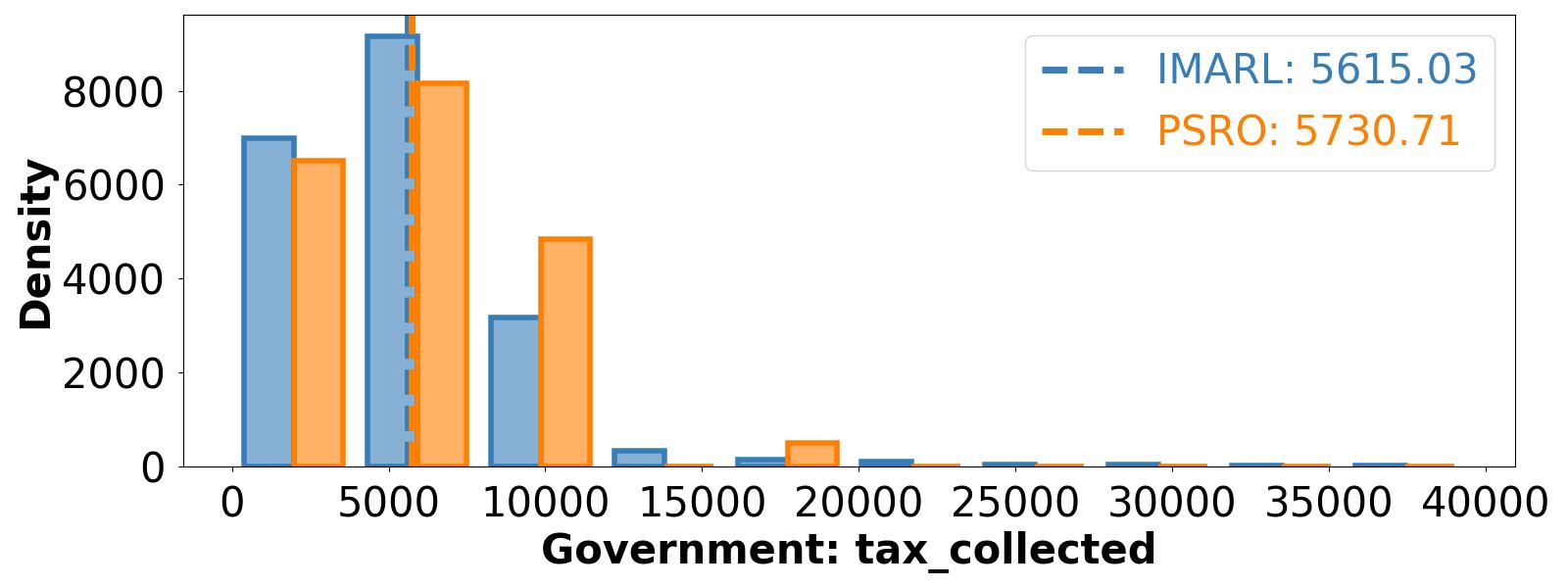}
    \caption{Tax rate (left) and total tax collected (right) by the government with IMARL and PSRO. The government is able to collect similar tax amounts from households while setting lower tax rates with PSRO due to higher labor hours.}
    \label{fig:gov_policy}
\end{figure}
\paragraph{Central Bank policy.}
Figure \ref{fig:cb_policy} compares interest rate set by the central bank (left) and inflation (right) under both schemes. 
With IMARL, the central bank utilizes all available rate options, whereas with PSRO, it tends to avoid the intermediate rates of 1.625\% and 4.375\%. 
Still, both methods successfully achieve the target inflation rate of $\pi^\star=1.02$ on average. 
But, we observe quarters with very high inflation with IMARL (see tails of blue histogram) unlike controlled inflation with PSRO. 
See the appendix for details on the temporal evolution of inflation and interest rates.

\paragraph{Government policy.}
Figure \ref{fig:gov_policy} compares tax rate (left) and total household tax collected (right) under both schemes. 
Although the government sets lower tax rates with PSRO than IMARL, the tax collected remains similar due increased labor hours with PSRO as in Figure \ref{fig:household_policy}.
PSRO agents learn an equilibrium where households work more while paying a smaller proportion of their income in taxes. 
In IMARL, high tax rates incentivize households to reduce labor to minimize tax payments, which in turn pushes the government to raise tax rates further to maintain credit redistribution.
In contrast, PSRO policies stabilize at an equilibrium where neither the households nor the government have an incentive to deviate.



\paragraph{Regret metrics.} 
Table \ref{tab:regret} shows the regret of strategies $\upsigma^{\textnormal{IMARL}}$ and $\upsigma^{\textnormal{PSRO}}$ against deviation set $\mathtt{\Sigma}=\mathtt{\Sigma}^{\textnormal{IMARL}}\cup\mathtt{\Sigma}^{\textnormal{PSRO}}$ across the four agent types, along with total regret. 
Since utility units differ across agent types, summing regret values to compute total regret may not be meaningful. So, we also report percentage regret as $\frac{\max_{\Tilde{\upsigma}_i\in\mathtt{\Sigma}}U_i\left(\Tilde{\upsigma}_i,\upsigma_{-i}\right)-U_i\left(\upsigma_i,\upsigma_{-i}\right)}{U_i\left(\upsigma_i,\upsigma_{-i}\right)}$, in parentheses next to absolute regret. Total percentage regret is got by dividing total regret by the total utility across agents for each scheme. While households experience zero regret under both schemes, PSRO achieves lower total regret compared to IMARL. Firms and government experience higher regret under IMARL, where firm 2 sets lower prices and higher wages, and the government collects slightly less in taxes. Both schemes result in low total percentage regret, with PSRO achieving near-zero total percentage regret.

\paragraph{Discussion.}
With PSRO, we observe a larger distinction between heterogeneous households: lower-skilled household 2 works more and consumes more than higher-skilled household 1. The distinction between heterogeneous firms is lower, with prices and wages getting closer while maintaining a small price difference for competitive advantage through households' consumption preferences. The central bank achieves target inflation with a smaller (but more disparate) set of rate options, and the government collects similar tax revenue with lower rates, driven by increased household labor compared to IMARL.
MARL is more challenging than single-agent RL due to environment non-stationarity from multiple evolving agents. While MARL policies are sensitive to the training scheme \cite{bowling2000analysis,bowling2002multiagent}, IMARL and PSRO also have different objectives.
IMARL focuses on concurrently finding policies that maximize rewards without considering equilibrium concepts. It doesn't address how a single agent might adjust its policy if those of other agents were fixed. However, PSRO with Nash equilibrium meta-strategies, aims to produce a set of policies where no agent has an incentive to change their policy if all others follow their PSRO-derived policies. 
To quantitatively compare policies from both schemes, we compute absolute regret and percentage regret that reflect utility loss when agents are restricted to one policy over another. We see that PSRO achieves lower total regret, indicating that agents following PSRO policies are less incentivized to change their policy, even with access to IMARL policies.

\section{Conclusion}
We aim to expand the use of agent-based modeling and reinforcement learning in economics by incorporating a multi-agent reinforcement learning (MARL) scheme grounded in equilibrium concepts. 
We introduce a highly configurable multi-agent economic simulator that allows users to define the number, types, and heterogeneity of objective-maximizing economic agents, enabling the simulation of various counterfactual scenarios. 
By implementing Policy Space Response Oracles (PSRO), a method combining empirical game-solving with deep RL, we learn policies for heterogeneous economic agents that form part of a Nash equilibrium, linking MARL to economic equilibria. 
Our comparison of PSRO and independent MARL in a scenario involving household skills and labor preferences underscores PSRO's strengths in finding policies with lower total regret. 
While analytically quantifying equilibria in economic systems with diverse, heterogeneous agents is challenging, we hope this framework with empirical game-solving broadens the use of agent-based modeling in economics, offering policymakers a powerful tool to explore policy outcomes in simulated economies \cite{farmer2009economy}.

\section*{Disclaimer}
This paper was prepared for informational purposes in part by the Artificial Intelligence Research group of JPMorgan Chase \& Co. and its affiliates (``JP Morgan'') and is not a product of the Research Department of JP Morgan. JP Morgan makes no representation and warranty whatsoever and disclaims all liability, for the completeness, accuracy or reliability of the information contained herein. This document is not intended as investment research or investment advice, or a recommendation, offer or solicitation for the purchase or sale of any security, financial instrument, financial product or service, or to be used in any way for evaluating the merits of participating in any transaction, and shall not constitute a solicitation under any jurisdiction or to any person, if such solicitation under such jurisdiction or to such person would be unlawful.

\bibliography{aaai25}

\begin{thebibliography}{76}
\providecommand{\natexlab}[1]{#1}

\bibitem[{Adler et~al.(2005)Adler, Satapathy, Manikonda, Bowles, and Blue}]{adler2005multi}
Adler, J.~L.; Satapathy, G.; Manikonda, V.; Bowles, B.; and Blue, V.~J. 2005.
\newblock A multi-agent approach to cooperative traffic management and route guidance.
\newblock \emph{Transportation Research Part B: Methodological}, 39(4): 297--318.

\bibitem[{Amrouni et~al.(2021)Amrouni, Moulin, Vann, Vyetrenko, Balch, and Veloso}]{amrouni2021abides}
Amrouni, S.; Moulin, A.; Vann, J.; Vyetrenko, S.; Balch, T.; and Veloso, M. 2021.
\newblock ABIDES-gym: gym environments for multi-agent discrete event simulation and application to financial markets.
\newblock In \emph{Proceedings of the Second ACM International Conference on AI in Finance}, 1--9.

\bibitem[{Arthur(2021)}]{arthur2021foundations}
Arthur, W.~B. 2021.
\newblock Foundations of complexity economics.
\newblock \emph{Nature Reviews Physics}, 3(2): 136--145.

\bibitem[{Atashbar and Shi(2023)}]{atashbar2023ai}
Atashbar, T.; and Shi, R.~A. 2023.
\newblock \emph{AI and macroeconomic modeling: Deep reinforcement learning in an RBC model}.
\newblock International Monetary Fund.

\bibitem[{Bowling and Veloso(2000)}]{bowling2000analysis}
Bowling, M.; and Veloso, M. 2000.
\newblock \emph{An analysis of stochastic game theory for multiagent reinforcement learning}.
\newblock Citeseer.

\bibitem[{Bowling and Veloso(2002)}]{bowling2002multiagent}
Bowling, M.; and Veloso, M. 2002.
\newblock Multiagent learning using a variable learning rate.
\newblock \emph{Artificial intelligence}, 136(2): 215--250.

\bibitem[{Brusatin et~al.(2024)Brusatin, Padoan, Coletta, Gatti, and Glielmo}]{brusatin2024simulating}
Brusatin, S.; Padoan, T.; Coletta, A.; Gatti, D.~D.; and Glielmo, A. 2024.
\newblock Simulating the economic impact of rationality through reinforcement learning and agent-based modelling.
\newblock \emph{arXiv preprint arXiv:2405.02161}.

\bibitem[{Busoniu, Babuska, and De~Schutter(2008)}]{busoniu2008comprehensive}
Busoniu, L.; Babuska, R.; and De~Schutter, B. 2008.
\newblock A comprehensive survey of multiagent reinforcement learning.
\newblock \emph{IEEE Transactions on Systems, Man, and Cybernetics}, 38(2): 156--172.

\bibitem[{Byrd, Hybinette, and Balch(2019)}]{byrd2019abides}
Byrd, D.; Hybinette, M.; and Balch, T.~H. 2019.
\newblock Abides: Towards high-fidelity market simulation for ai research.
\newblock \emph{arXiv preprint arXiv:1904.12066}.

\bibitem[{Chen et~al.(2021)Chen, Joseph, Kumhof, Pan, Shi, and Zhou}]{chen2021deep}
Chen, M.; Joseph, A.; Kumhof, M.; Pan, X.; Shi, R.; and Zhou, X. 2021.
\newblock Deep reinforcement learning in a monetary model.
\newblock \emph{arXiv preprint arXiv:2104.09368}.

\bibitem[{Christiano, Eichenbaum, and Evans(2005)}]{christiano2005nominal}
Christiano, L.~J.; Eichenbaum, M.; and Evans, C.~L. 2005.
\newblock Nominal rigidities and the dynamic effects of a shock to monetary policy.
\newblock \emph{Journal of political Economy}, 113(1): 1--45.

\bibitem[{Claus and Boutilier(1998)}]{claus1998dynamics}
Claus, C.; and Boutilier, C. 1998.
\newblock The dynamics of reinforcement learning in cooperative multiagent systems.
\newblock \emph{AAAI/IAAI}, 1998(746-752): 2.

\bibitem[{Cobb and Douglas(1928)}]{cobb1928theory}
Cobb, C.~W.; and Douglas, P.~H. 1928.
\newblock A theory of production.
\newblock \emph{American Economic Association}.

\bibitem[{Curry et~al.(2022)Curry, Trott, Phade, Bai, Zheng et~al.}]{curry2022analyzing}
Curry, M.; Trott, A.; Phade, S.; Bai, Y.; Zheng, S.; et~al. 2022.
\newblock Analyzing Micro-Founded General Equilibrium Models with Many Agents using Deep Reinforcement Learning.
\newblock Technical report.

\bibitem[{Dawid et~al.(2016)Dawid, Harting, van~der Hoog, and Neugart}]{dawid2016heterogeneous}
Dawid, H.; Harting, P.; van~der Hoog, S.; and Neugart, M. 2016.
\newblock A heterogeneous agent macroeconomic model for policy evaluation: Improving transparency and reproducibility.

\bibitem[{Deissenberg, Van Der~Hoog, and Dawid(2008)}]{deissenberg2008eurace}
Deissenberg, C.; Van Der~Hoog, S.; and Dawid, H. 2008.
\newblock EURACE: A massively parallel agent-based model of the European economy.
\newblock \emph{Applied mathematics and computation}, 204(2): 541--552.

\bibitem[{Del~Negro et~al.(2015)Del~Negro, Giannoni, Li, Moszkowski, and Smith}]{frbny_dsge_jl}
Del~Negro, M.; Giannoni, M.; Li, P.; Moszkowski, E.; and Smith, M. 2015.
\newblock New York Fed DSGE Model (Version 1002).
\newblock \url{https://github.com/FRBNY-DSGE/DSGE.jl}.

\bibitem[{Del~Negro, Giannoni, and Schorfheide(2015)}]{negro2015}
Del~Negro, M.; Giannoni, M.~P.; and Schorfheide, F. 2015.
\newblock Inflation in the Great Recession and New Keynesian Models.
\newblock \emph{American Economic Journal: Macroeconomics}, 7(1): 168--96.

\bibitem[{Del~Negro and Schorfheide(2013)}]{del2013dsge}
Del~Negro, M.; and Schorfheide, F. 2013.
\newblock DSGE model-based forecasting.
\newblock In \emph{Handbook of economic forecasting}, volume~2, 57--140. Elsevier.

\bibitem[{Dong, Dwarakanath, and Vyetrenko(2023)}]{dong2023analyzing}
Dong, J.; Dwarakanath, K.; and Vyetrenko, S. 2023.
\newblock Analyzing the Impact of Tax Credits on Households in Simulated Economic Systems with Learning Agents.
\newblock \emph{arXiv preprint arXiv:2311.17252}.

\bibitem[{Dosi et~al.(2015)Dosi, Fagiolo, Napoletano, Roventini, and Treibich}]{dosi2015fiscal}
Dosi, G.; Fagiolo, G.; Napoletano, M.; Roventini, A.; and Treibich, T. 2015.
\newblock Fiscal and monetary policies in complex evolving economies.
\newblock \emph{Journal of Economic Dynamics and Control}, 52: 166--189.

\bibitem[{Dosi, Fagiolo, and Roventini(2006)}]{dosi2006evolutionary}
Dosi, G.; Fagiolo, G.; and Roventini, A. 2006.
\newblock An evolutionary model of endogenous business cycles.
\newblock \emph{Computational Economics}, 27: 3--34.

\bibitem[{Dosi et~al.(2017)Dosi, Napoletano, Roventini, and Treibich}]{dosi2017micro}
Dosi, G.; Napoletano, M.; Roventini, A.; and Treibich, T. 2017.
\newblock Micro and macro policies in the Keynes+ Schumpeter evolutionary models.
\newblock \emph{Journal of Evolutionary Economics}, 27: 63--90.

\bibitem[{Dwarakanath et~al.(2024)Dwarakanath, Vyetrenko, Tavallali, and Balch}]{dwarakanath2024abides}
Dwarakanath, K.; Vyetrenko, S.; Tavallali, P.; and Balch, T. 2024.
\newblock ABIDES-Economist: Agent-Based Simulation of Economic Systems with Learning Agents.
\newblock \emph{arXiv preprint arXiv:2402.09563}.

\bibitem[{Evans and Honkapohja(2005)}]{evans2005policy}
Evans, G.~W.; and Honkapohja, S. 2005.
\newblock Policy interaction, expectations and the liquidity trap.
\newblock \emph{Review of Economic Dynamics}, 8(2): 303--323.

\bibitem[{Farmer and Foley(2009)}]{farmer2009economy}
Farmer, J.~D.; and Foley, D. 2009.
\newblock The economy needs agent-based modelling.
\newblock \emph{Nature}, 460(7256): 685--686.

\bibitem[{{Federal Reserve Board}(2023)}]{fr_rate}
{Federal Reserve Board}. 2023.
\newblock Selected Interest Rates (Daily).
\newblock \url{https://www.federalreserve.gov/releases/h15/}.

\bibitem[{Fudenberg and Levine(1998)}]{fudenberg1998theory}
Fudenberg, D.; and Levine, D.~K. 1998.
\newblock \emph{The theory of learning in games}, volume~2.
\newblock MIT press.

\bibitem[{Gatti et~al.(2014)Gatti, Cavalin, Neto, Pinhanez, dos Santos, Gribel, and Appel}]{gatti2014large}
Gatti, M.; Cavalin, P.; Neto, S.~B.; Pinhanez, C.; dos Santos, C.; Gribel, D.; and Appel, A.~P. 2014.
\newblock Large-scale multi-agent-based modeling and simulation of microblogging-based online social network.
\newblock In \emph{Multi-Agent-Based Simulation XIV: International Workshop}, 17--33.

\bibitem[{Hahn(1973)}]{hahn1973notion}
Hahn, F.~H. 1973.
\newblock \emph{On the notion of equilibrium in economics: an inaugural lecture}.
\newblock Cambridge University Press.

\bibitem[{Haldane and Turrell(2019)}]{haldane2019drawing}
Haldane, A.~G.; and Turrell, A.~E. 2019.
\newblock Drawing on different disciplines: macroeconomic agent-based models.
\newblock \emph{Journal of Evolutionary Economics}, 29: 39--66.

\bibitem[{Hamill and Gilbert(2015)}]{hamill2015agent}
Hamill, L.; and Gilbert, N. 2015.
\newblock \emph{Agent-based modelling in economics}.
\newblock John Wiley \& Sons.

\bibitem[{Hildenbrand(1983)}]{hildenbrand1983law}
Hildenbrand, W. 1983.
\newblock On the ``law of Demand".
\newblock \emph{Econometrica: Journal of the Econometric Society}, 997--1019.

\bibitem[{Hill, Bardoscia, and Turrell(2021)}]{hill2021solving}
Hill, E.; Bardoscia, M.; and Turrell, A. 2021.
\newblock Solving heterogeneous general equilibrium economic models with deep reinforcement learning.
\newblock \emph{arXiv preprint arXiv:2103.16977}.

\bibitem[{Hinterlang and T{\"a}nzer(2021)}]{hinterlang2021optimal}
Hinterlang, N.; and T{\"a}nzer, A. 2021.
\newblock Optimal monetary policy using reinforcement learning.
\newblock Technical report, Deutsche Bundesbank Discussion Paper.

\bibitem[{Hu and Wellman(1998)}]{hu1998multiagent}
Hu, J.; and Wellman, M.~P. 1998.
\newblock Multiagent Reinforcement Learning: Theoretical Framework and an Algorithm.
\newblock In \emph{Proceedings of the Fifteenth International Conference on Machine Learning}, 242--250.

\bibitem[{Hu and Wellman(2003)}]{hu2003nash}
Hu, J.; and Wellman, M.~P. 2003.
\newblock Nash Q-learning for general-sum stochastic games.
\newblock \emph{Journal of machine learning research}, 4(Nov): 1039--1069.

\bibitem[{{Internal Revenue Service}(2022)}]{irs_tax}
{Internal Revenue Service}. 2022.
\newblock 2022 Tax Rate Schedules.
\newblock \url{https://www.irs.gov/media/166986}.

\bibitem[{Kaelbling, Littman, and Moore(1996)}]{kaelbling1996reinforcement}
Kaelbling, L.~P.; Littman, M.~L.; and Moore, A.~W. 1996.
\newblock Reinforcement Learning: A Survey.
\newblock \emph{Journal of Artificial Intelligence Research}, 4: 237--285.

\bibitem[{Knight and Campbell(2018)}]{nashpy}
Knight, V.; and Campbell, J. 2018.
\newblock Nashpy: A Python library for the computation of Nash equilibria.
\newblock \emph{Journal of Open Source Software}, 3(30): 904.

\bibitem[{Koster et~al.(2022)Koster, Balaguer, Tacchetti, Weinstein, Zhu, Hauser, Williams, Campbell-Gillingham, Thacker, Botvinick et~al.}]{koster2022human}
Koster, R.; Balaguer, J.; Tacchetti, A.; Weinstein, A.; Zhu, T.; Hauser, O.; Williams, D.; Campbell-Gillingham, L.; Thacker, P.; Botvinick, M.; et~al. 2022.
\newblock Human-centred mechanism design with Democratic AI.
\newblock \emph{Nature Human Behaviour}, 6(10): 1398--1407.

\bibitem[{Krusell and Smith(1998)}]{krusell1998income}
Krusell, P.; and Smith, A.~A., Jr. 1998.
\newblock Income and wealth heterogeneity in the macroeconomy.
\newblock \emph{Journal of political Economy}, 106(5): 867--896.

\bibitem[{Kydland and Prescott(1982)}]{kydland1982time}
Kydland, F.~E.; and Prescott, E.~C. 1982.
\newblock Time to build and aggregate fluctuations.
\newblock \emph{Econometrica: Journal of the Econometric Society}, 1345--1370.

\bibitem[{Lanctot et~al.(2017)Lanctot, Zambaldi, Gruslys, Lazaridou, Tuyls, P{\'e}rolat, Silver, and Graepel}]{lanctot2017unified}
Lanctot, M.; Zambaldi, V.; Gruslys, A.; Lazaridou, A.; Tuyls, K.; P{\'e}rolat, J.; Silver, D.; and Graepel, T. 2017.
\newblock A unified game-theoretic approach to multiagent reinforcement learning.
\newblock \emph{Advances in neural information processing systems}, 30.

\bibitem[{Liang et~al.(2018)Liang, Liaw, Nishihara, Moritz, Fox, Goldberg, Gonzalez, Jordan, and Stoica}]{rllib}
Liang, E.; Liaw, R.; Nishihara, R.; Moritz, P.; Fox, R.; Goldberg, K.; Gonzalez, J.; Jordan, M.; and Stoica, I. 2018.
\newblock RLlib: Abstractions for Distributed Reinforcement Learning.
\newblock In \emph{International Conference on Machine Learning}.

\bibitem[{Littman(1994)}]{littman1994markov}
Littman, M.~L. 1994.
\newblock Markov games as a framework for multi-agent reinforcement learning.
\newblock In \emph{Machine learning proceedings}, 157--163.

\bibitem[{Lundberg and Lee(2017)}]{shap}
Lundberg, S.~M.; and Lee, S. 2017.
\newblock A Unified Approach to Interpreting Model Predictions.
\newblock In Guyon, I.; Luxburg, U.~V.; Bengio, S.; Wallach, H.; Fergus, R.; Vishwanathan, S.; and Garnett, R., eds., \emph{Advances in Neural Information Processing Systems 30}, 4765--4774. Long Beach: Curran Associates, Inc.

\bibitem[{Macal and North(2005)}]{macal2005tutorial}
Macal, C.~M.; and North, M.~J. 2005.
\newblock Tutorial on agent-based modeling and simulation.
\newblock In \emph{Proceedings of the Winter Simulation Conference}.

\bibitem[{McMahan, Gordon, and Blum(2003)}]{mcmahan2003planning}
McMahan, H.~B.; Gordon, G.~J.; and Blum, A. 2003.
\newblock Planning in the presence of cost functions controlled by an adversary.
\newblock In \emph{Proceedings of the 20th International Conference on Machine Learning (ICML-03)}, 536--543.

\bibitem[{Mi et~al.(2023)Mi, Xia, Song, Zhang, Zhu, and Wang}]{mi2023taxai}
Mi, Q.; Xia, S.; Song, Y.; Zhang, H.; Zhu, S.; and Wang, J. 2023.
\newblock TaxAI: A Dynamic Economic Simulator and Benchmark for Multi-Agent Reinforcement Learning.
\newblock \emph{arXiv preprint arXiv:2309.16307}.

\bibitem[{Park et~al.(2023)Park, O'Brien, Cai, Morris, Liang, and Bernstein}]{park2023generative}
Park, J.~S.; O'Brien, J.~C.; Cai, C.~J.; Morris, M.~R.; Liang, P.; and Bernstein, M.~S. 2023.
\newblock Generative agents: Interactive simulacra of human behavior.
\newblock \emph{arXiv preprint arXiv:2304.03442}.

\bibitem[{Savani and Turocy(2023)}]{gambit}
Savani, R.; and Turocy, T.~L. 2023.
\newblock Gambit: The package for computation in game theory, Version 16.0.2.
\newblock \url{http://www.gambit-project.org}.

\bibitem[{Schulman et~al.(2017)Schulman, Wolski, Dhariwal, Radford, and Klimov}]{ppo}
Schulman, J.; Wolski, F.; Dhariwal, P.; Radford, A.; and Klimov, O. 2017.
\newblock Proximal policy optimization algorithms.
\newblock \emph{arXiv preprint arXiv:1707.06347}.

\bibitem[{Schvartzman and Wellman(2009)}]{schvartzman2009stronger}
Schvartzman, L.~J.; and Wellman, M.~P. 2009.
\newblock Stronger CDA strategies through empirical game-theoretic analysis and reinforcement learning.
\newblock In \emph{Proceedings of The 8th International Conference on Autonomous Agents and Multiagent Systems-Volume 1}, 249--256.

\bibitem[{Smith, Anthony, and Wellman(2020)}]{smith2020iterative}
Smith, M.; Anthony, T.; and Wellman, M. 2020.
\newblock Iterative Empirical Game Solving via Single Policy Best Response.
\newblock In \emph{International Conference on Learning Representations}.

\bibitem[{Srbljinovi{\'c} and {\v{S}}kunca(2003)}]{srbljinovic2003introduction}
Srbljinovi{\'c}, A.; and {\v{S}}kunca, O. 2003.
\newblock An introduction to agent based modelling and simulation of social processes.
\newblock \emph{Interdisciplinary Description of Complex Systems}, 1(1-2): 1--8.

\bibitem[{{Statista}(2023)}]{statista}
{Statista}. 2023.
\newblock Per capita consumption of the bread and cereal products in the United States from 2017 to 2027.
\newblock \url{https://www.statista.com/forecasts/1374278/size-of-the-bread-and-cereal-product-market-in-the-united-states}.

\bibitem[{Stiglitz(2018)}]{stiglitz2018modern}
Stiglitz, J.~E. 2018.
\newblock Where modern macroeconomics went wrong.
\newblock \emph{Oxford Review of Economic Policy}, 34(1-2): 70--106.

\bibitem[{Svensson(2020)}]{svensson2020monetary}
Svensson, L.~E. 2020.
\newblock Monetary policy strategies for the Federal Reserve.
\newblock Technical report, National Bureau of Economic Research.

\bibitem[{Tan(1993)}]{tan1993multi}
Tan, M. 1993.
\newblock Multi-agent reinforcement learning: Independent vs. cooperative agents.
\newblock In \emph{Proceedings of the tenth international conference on machine learning}, 330--337.

\bibitem[{Taylor and Williams(2010)}]{taylor2010simple}
Taylor, J.~B.; and Williams, J.~C. 2010.
\newblock Simple and robust rules for monetary policy.
\newblock In \emph{Handbook of monetary economics}, volume~3, 829--859. Elsevier.

\bibitem[{Tesfatsion and Judd(2006)}]{tesfatsion2006handbook}
Tesfatsion, L.; and Judd, K.~L. 2006.
\newblock \emph{Handbook of computational economics: agent-based computational economics}.
\newblock Elsevier.

\bibitem[{Trott et~al.(2021)Trott, Srinivasa, van~der Wal, Haneuse, and Zheng}]{trott2021building}
Trott, A.; Srinivasa, S.; van~der Wal, D.; Haneuse, S.; and Zheng, S. 2021.
\newblock Building a foundation for data-driven, interpretable, and robust policy design using the ai economist.
\newblock \emph{arXiv preprint arXiv:2108.02904}.

\bibitem[{{U.S. Bureau of Labor Statistics}(2023{\natexlab{a}})}]{bls_avg_wage}
{U.S. Bureau of Labor Statistics}. 2023{\natexlab{a}}.
\newblock Average hourly and weekly earnings of all employees on private nonfarm payrolls by industry sector, seasonally adjusted.
\newblock \url{https://www.bls.gov/news.release/empsit.t19.htm}.

\bibitem[{{U.S. Bureau of Labor Statistics}(2023{\natexlab{b}})}]{bls_price}
{U.S. Bureau of Labor Statistics}. 2023{\natexlab{b}}.
\newblock Average price data (in U.S. dollars), selected items.
\newblock \url{https://www.bls.gov/charts/consumer-price-index/consumer-price-index-average-price-data.htm}.

\bibitem[{{U.S. Bureau of Labor Statistics}(2023{\natexlab{c}})}]{bls}
{U.S. Bureau of Labor Statistics}. 2023{\natexlab{c}}.
\newblock Employment Situation.
\newblock \url{https://www.bls.gov/news.release/empsit.toc.htm}.

\bibitem[{USA.gov(2023)}]{usa_gov_min_wage}
USA.gov. 2023.
\newblock Minimum wage.
\newblock \url{https://www.usa.gov/minimum-wage}.

\bibitem[{Von~Neumann(1928)}]{von1928theory}
Von~Neumann, J. 1928.
\newblock On the Theory of Games of Strategy.
\newblock \emph{Mathematische Annalen}, 100: 295--320.

\bibitem[{Von~Neumann and Morgenstern(2007)}]{von2007theory}
Von~Neumann, J.; and Morgenstern, O. 2007.
\newblock \emph{Theory of games and economic behavior (60th Anniversary Commemorative Edition)}.
\newblock Princeton university press.

\bibitem[{Vorotnikov et~al.(2018)Vorotnikov, Ermishin, Nazarova, and Yuschenko}]{vorotnikov2018multi}
Vorotnikov, S.; Ermishin, K.; Nazarova, A.; and Yuschenko, A. 2018.
\newblock Multi-agent robotic systems in collaborative robotics.
\newblock In \emph{Interactive Collaborative Robotics: Third International Conference}, 270--279.

\bibitem[{Walsh et~al.(2002)Walsh, Das, Tesauro, and Kephart}]{walsh2002analyzing}
Walsh, W.~E.; Das, R.; Tesauro, G.; and Kephart, J.~O. 2002.
\newblock Analyzing complex strategic interactions in multi-agent systems.
\newblock In \emph{AAAI-02 Workshop on Game-Theoretic and Decision-Theoretic Agents}, 109--118.

\bibitem[{Wellman(2006)}]{wellman2006methods}
Wellman, M.~P. 2006.
\newblock Methods for empirical game-theoretic analysis.
\newblock In \emph{AAAI}, volume 980, 1552--1556.

\bibitem[{Wellman(2020)}]{wellman2020economic}
Wellman, M.~P. 2020.
\newblock Economic reasoning from simulation-based game models.
\newblock \emph{{\OE}conomia. History, Methodology, Philosophy}, (10-2): 257--278.

\bibitem[{Woodford(2009)}]{woodford2009convergence}
Woodford, M. 2009.
\newblock Convergence in macroeconomics: elements of the new synthesis.
\newblock \emph{American economic journal}, 1(1): 267--279.

\bibitem[{Wright, Wang, and Wellman(2019)}]{iter_deep_rl_mason}
Wright, M.; Wang, Y.; and Wellman, M.~P. 2019.
\newblock Iterated Deep Reinforcement Learning in Games: History-Aware Training for Improved Stability.
\newblock In \emph{Proceedings of the 2019 ACM Conference on Economics and Computation}, EC '19, 617–636. New York, NY, USA: Association for Computing Machinery.

\bibitem[{Zheng et~al.(2022)Zheng, Trott, Srinivasa, Parkes, and Socher}]{zheng2022}
Zheng, S.; Trott, A.; Srinivasa, S.; Parkes, D.~C.; and Socher, R. 2022.
\newblock The {AI} Economist: {T}axation policy design via two-level deep multiagent reinforcement learning.
\newblock \emph{Science Advances}, 8(18): 2607.

\end{thebibliography}

\appendix
\section*{Calibration and Realism}\label{appsec:calibration}

Table \ref{tab:calibration} provides details on the values and sources for default agent parameters in our simulator. 
Default agent parameters are replaced by otherwise specified values when modeling heterogeneity. Agent action spaces comprise a uniform grid of values around the default values in \textbf{bold}, while adhering to any minimum value constraints.
\begin{table*}[tb]
    \centering
    \begin{tabular}{lllll}\toprule
        Agent & Variable Type & Notation & Value & Source \\\midrule
        Household $i$ & Parameter & $\omega_{ij}$ & $1.00$ & \\
        & & $\gamma_i$ & $0.33$ & \cite{chen2021deep}\\
        & & $\nu_i$ & $0.50$ & \cite{chen2021deep}\\
        & & $\mu_i$ & $0.10$ & \cite{chen2021deep}\\
        & & $\beta_{i,\mathbf{H}}$ & $0.99$ & \cite{chen2021deep}\\
        & Action & $n_{t,ij}$ & $\lbrace0,240,\textbf{480},720,960\rbrace$ & 40 hours per week \\
        & & & & $\approx$ 480 hours per quarter (12 weeks)\\
        & & $c^{req}_{t,ij}$& $\lbrace0,6,\textbf{12},18,24\rbrace$ & Per capita consumption of 1lb\\
        & & & & of bread per week \cite{statista}.\\
        \midrule
        Firm $j$ & Parameter & $\rho_j,\bar{\varepsilon}_{j},\sigma_{j}$ & $0.97,0.00,0.10$ & \cite{hill2021solving}\\
        & & $\alpha_j$ & $\frac{2}{3}$ & \cite{hill2021solving}\\
        & & $\chi_j$ & $0.10$ & \\
        & & $\beta_{j,\mathbf{F}}$ & $0.99$ & \\
        & Action & $w_{t,j}$ & $\lbrace7.25,19.65,\textbf{32.06},44.46,56.87\rbrace$ & Minimum wage \cite{usa_gov_min_wage} and \\
        & & & & average hourly earnings in May 2022\\
        & & & & \cite{bls_avg_wage}.\\
        & & $p_{t,j}$ & $\lbrace188,255,\textbf{322},389,456\rbrace$ & Price of bread/lb in May 2022\\
        & & & & \cite{bls_price}\\
        & & & & multiplied by 200 consumable goods.\\
        \midrule
        Central Bank & Parameter & $\pi^\star$ & $1.02$ & \cite{svensson2020monetary,hinterlang2021optimal}\\
        & & $\lambda$ & $0.25$ & \cite{svensson2020monetary}\\
        & & $\beta_{\mathbf{CB}}$ & $0.99$ & \cite{hinterlang2021optimal}\\
        & Action & $r_t$ & $\lbrace0.00250,0.01625,\textbf{0.03},0.04375,0.05750\rbrace$ & Federal funds rate\\
        & & & & \cite{fr_rate}\\
        \midrule
        Government & Parameter & $\xi$ & $0.10$ \\
        & & $\beta_{\mathbf{G}}$ & $0.99$ & \\
        & Action & $\tau_t$ & $\lbrace0.1000,0.1675,\textbf{0.2350},0.3025,0.3700\rbrace$ & Lowest to highest tax brackets in 2022\\
        & & & & \cite{irs_tax}\\
        & & $f_{t,i}$ & $\lbrace1,2,\textbf{3},4,5\rbrace$ then, normalized by  $\sum_kf_{t,k}$\\
        \bottomrule
    \end{tabular}
    \caption{Default agent parameters and agent action spaces in our simulator. 
    \textbf{Bold-faced} values represent default actions.}
    \label{tab:calibration}
\end{table*}
\begin{table*}[h!]
    \centering
    \begin{tabular}{lll}\toprule
        Agent & Reward & Normalized reward \\
        \midrule
        Household $i$ & $\sum_{j}u(c_{t,ij},n_{t,ij},m_{t+1,i};\gamma_i,\nu_i,\mu_i)$ & $\sum_{j}u(c_{t,ij},\frac{n_{t,ij}}{\bar{n}_i},\frac{m_{t+1,i}}{\left(\bar{n}_i\sum_j\bar{w}_j\right)\left(\frac{\sum_jp_{t,j}}{\sum_j1}\right)};\gamma_i,\nu_i,\mu_i)$\\
        \midrule
        Firm $j$ & $p_{t,j}\sum_{i}c_{t,ij}-w_{t,j}\sum_in_{t,ij}\omega_{ij}-\chi_jp_{t,j}Y_{t+1,j}$ & $\frac{p_{t,j}\sum_{i}c_{t,ij}}{\bar{p}_j\sum_i\bar{c}_i}-\frac{w_{t,j}\sum_in_{t,ij}}{\bar{w}_j\sum_i\bar{n}_i}\omega_{ij}-\chi_j\frac{p_{t,j}Y_{t+1,j}}{\bar{p}_j\exp(\bar\varepsilon_j+10\sigma_j)\sum_i\bar{n}_i}$\\
        \midrule
        Central Bank & $-\left(\pi_t-\pi^\star\right)^2+\lambda\left(\sum_jy_{t,j}\right)^2$ & $-\left(\pi_t-\pi^\star\right)^2+\lambda\left(\frac{\sum_jy_{t,j}}{\sum_j\bar{y}_j}\right)^2$ where $\bar{y}_j=\left(\sum_i\bar{n}_i\right)^{\alpha_j}$ \\
        \midrule
        Government & $\sum_il_{t,i}R_{t,i,\mathbf{H}}$ & $\sum_il_{t,i}\times\textnormal{Normalized reward of Household }i$\\
        \bottomrule
    \end{tabular}
    \caption{Normalization of agent rewards. Values for default labor hours $\bar{n}_i$, consumption $\Bar{c}_i$, price $\Bar{p}_j$ and wage $\bar{w}_j$ are given by the \textbf{bold-faced} values in Table \ref{tab:calibration}.}
    \label{tab:norm_reward}
\end{table*}

\section*{Verification of Stylized Facts}\label{appsec:stylized_facts}
\begin{figure}[tb]
    \centering
    \includegraphics[width=0.4\linewidth]{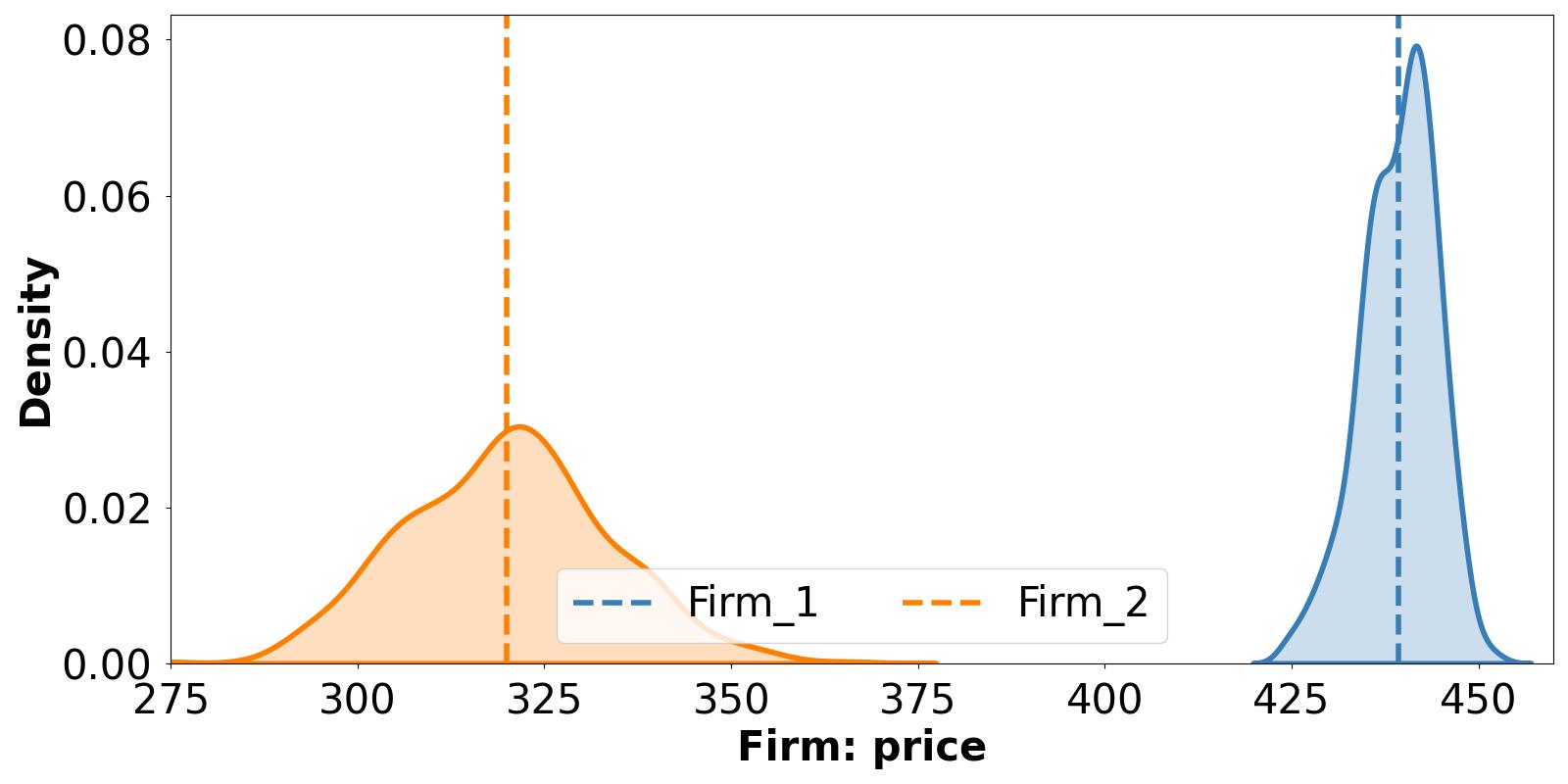}\hspace{0.05\linewidth}
    \includegraphics[width=0.4\linewidth]{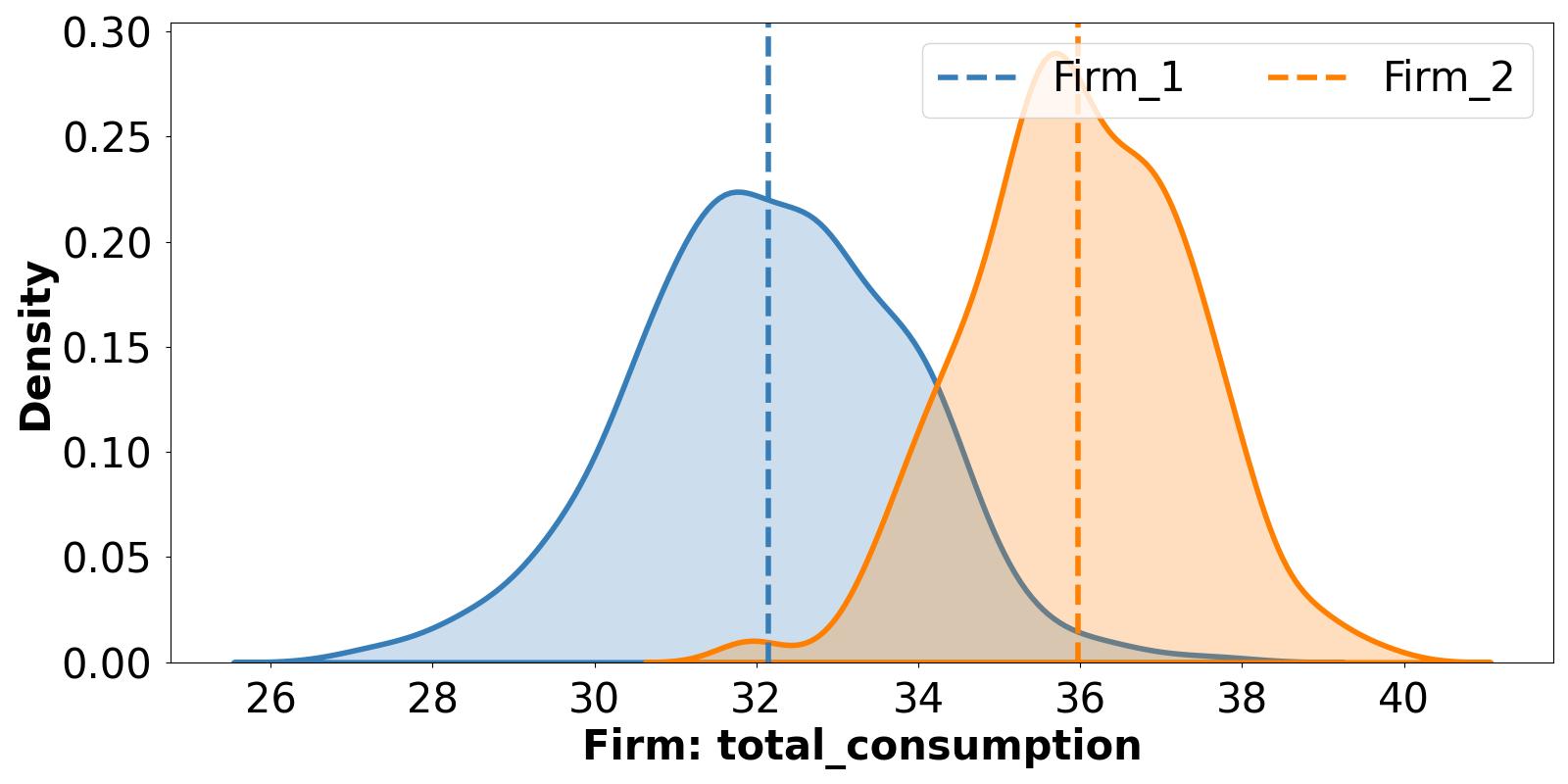}
    \caption{Prices set by firms (left) and household consumption (right) with IMARL. Observe that firm 1 that sets higher price receives lower consumption from households verifying the law of demand.}
    \label{fig:law_of_demand}
\end{figure}

To illustrate the ability of our simulator to reproduce economic stylized facts, we play out the agent policies learned with IMARL in test episodes where we observe conformity to the following stylized facts.
\paragraph{The Law of Demand.}The law of demand states that consumption of a good decreases as the price of the good increases given that other factors remain the same \cite{hildenbrand1983law}. We plot the prices set by both firms (left) alongside the total consumption across households per firm (right) in Figure \ref{fig:law_of_demand}. We plot the distribution across test episodes of average firm prices per episode. We observe that firm 1 that sets higher prices receives lower consumption.

\begin{figure}[tb]
    \centering
    \includegraphics[width=0.9\linewidth]{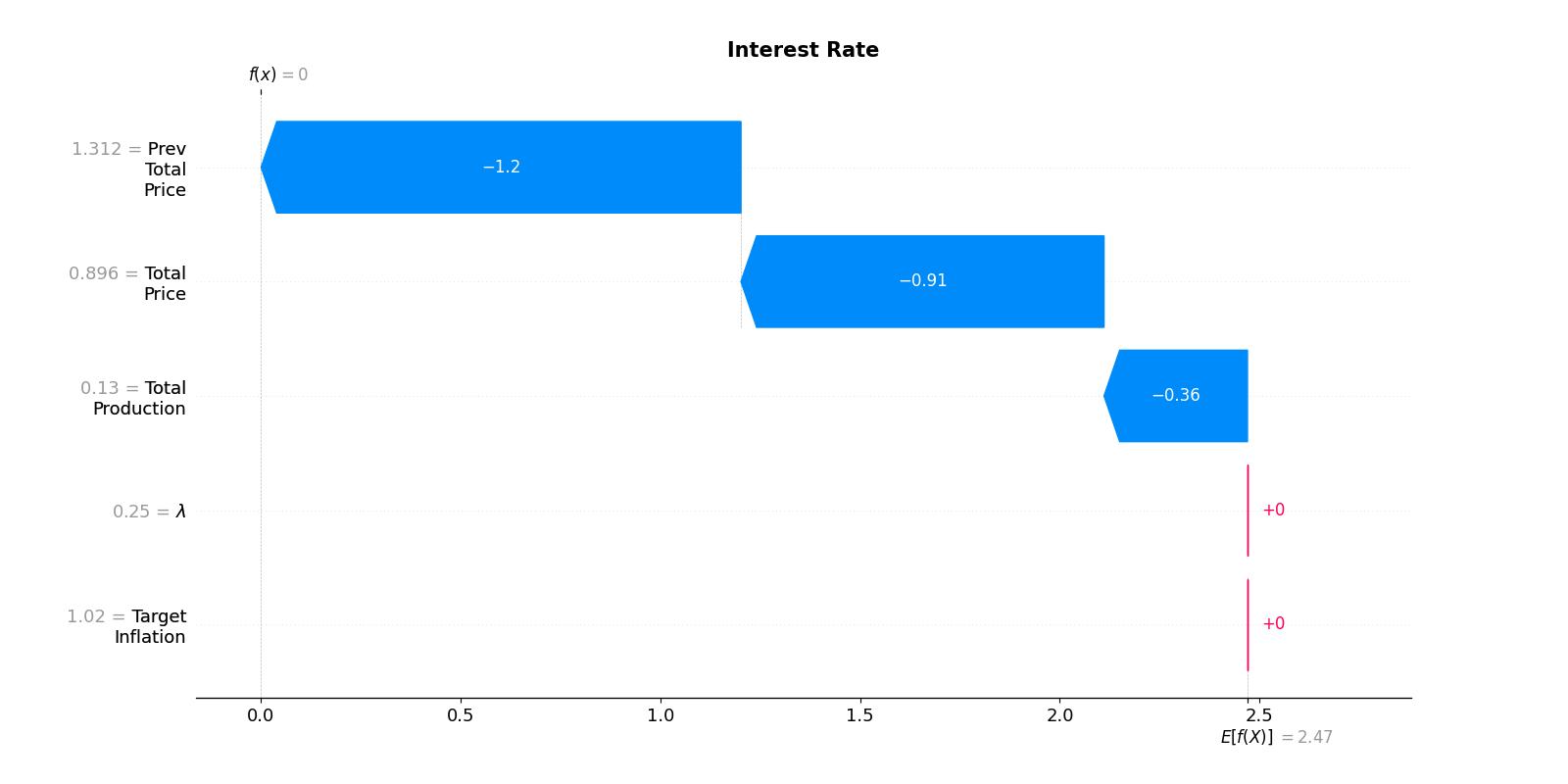}
    \caption{SHAP analysis of Central Bank policy that sets interest rate given observations. Notice that high previous price and relatively lower current price influence rate negatively alongside low production influencing rate negatively.}
    \label{fig:CB_waterfall}
\end{figure}

\paragraph{Positive impact of inflation on interest rate.}Standard monetary policy rules express the interest rate set by the Central Bank (CB) as an increasing function of inflation \cite{taylor2010simple}. Thus, the interest rate is raised in response to high inflation and, is lowered in response to low inflation. To study the impact of inflation on the learned CB policy, we perform an explainability analysis of the Proximal Policy Optimization policy network that takes in observations to give out CB action of interest rate. We use the tool called SHAP (for SHapley Additive exPlanations) to decompose the network output locally into a sum of effects attributed to each observation feature \cite{shap}. Figure \ref{fig:CB_waterfall} shows the impact of the CB's observation features on interest rate, sorted in decreasing order of their importances. The length of each bar corresponds to the magnitude of importance of the feature with red showing positive impact and blue showing negative impact. The feature names are preceded by their numerical values on the vertical axis. Figure \ref{fig:CB_waterfall} is interpreted as follows.
The previous total price across firms is the most impactful feature, followed by the current total price and then, the total production across firms. Observe that current total price is lower than that previously, indicating low inflation which impacts interest rate in a negative manner. This verifies the positive relationship between inflation and interest rate. In addition, low value for production impacts interest rate in a negative manner. This is because when production is low, the CB wants to increase production by pushing households to provide more labor. This is done by reducing interest rates so that households earn lower interest on their savings, causing them to provide labor so as to earn labor income.

\section*{Learning Details}\label{appsec:learning}
Each agent has continuous observation space and discrete action space as described in Table \ref{tab:calibration}. 
In order to ease learning, we normalize agent rewards as in Table \ref{tab:norm_reward}. Learning rates per agent type are set via a grid search over $\lbrace10^{-3},2\times10^{-3},5\times10^{-3},10^{-2}\rbrace$ for each learning scenario. The final learning rates are the first set that achieved improvement and convergence in training rewards for all agents.

\section*{Compute Details}
We parallelize training across agent types per PSRO epoch in line 4 of algorithm \ref{alg:psro}. Per agent type, the policy is updated sequentially from one RL episode to the next in line 5. And, PSRO epochs run in sequence in line 3. For line 11, we run 10 parallel simulations per joint policy to compute agent utilities as their average discounted cumulative rewards over those runs. This was run on an AWS ec2 instance of type c5.12xlarge, with Linux OS, 48 vCPUs and no GPU. Experiments were run using python 3.7.16 with gym0.22.0, ray1.7.0 and tensorflow2.8.0. The training time for 8 PSRO epochs with 100 RL episodes per epoch was about 7 hours. For comparison, IMARL training over 4000 episodes took under an hour on the same instance without any additional parallelization on our end. 
\section*{Additional Experimental Results}\label{appsec:expts}
\subsection*{Central Bank policy}
\begin{figure}[tb]
    \centering
    \includegraphics[width=0.4\linewidth]{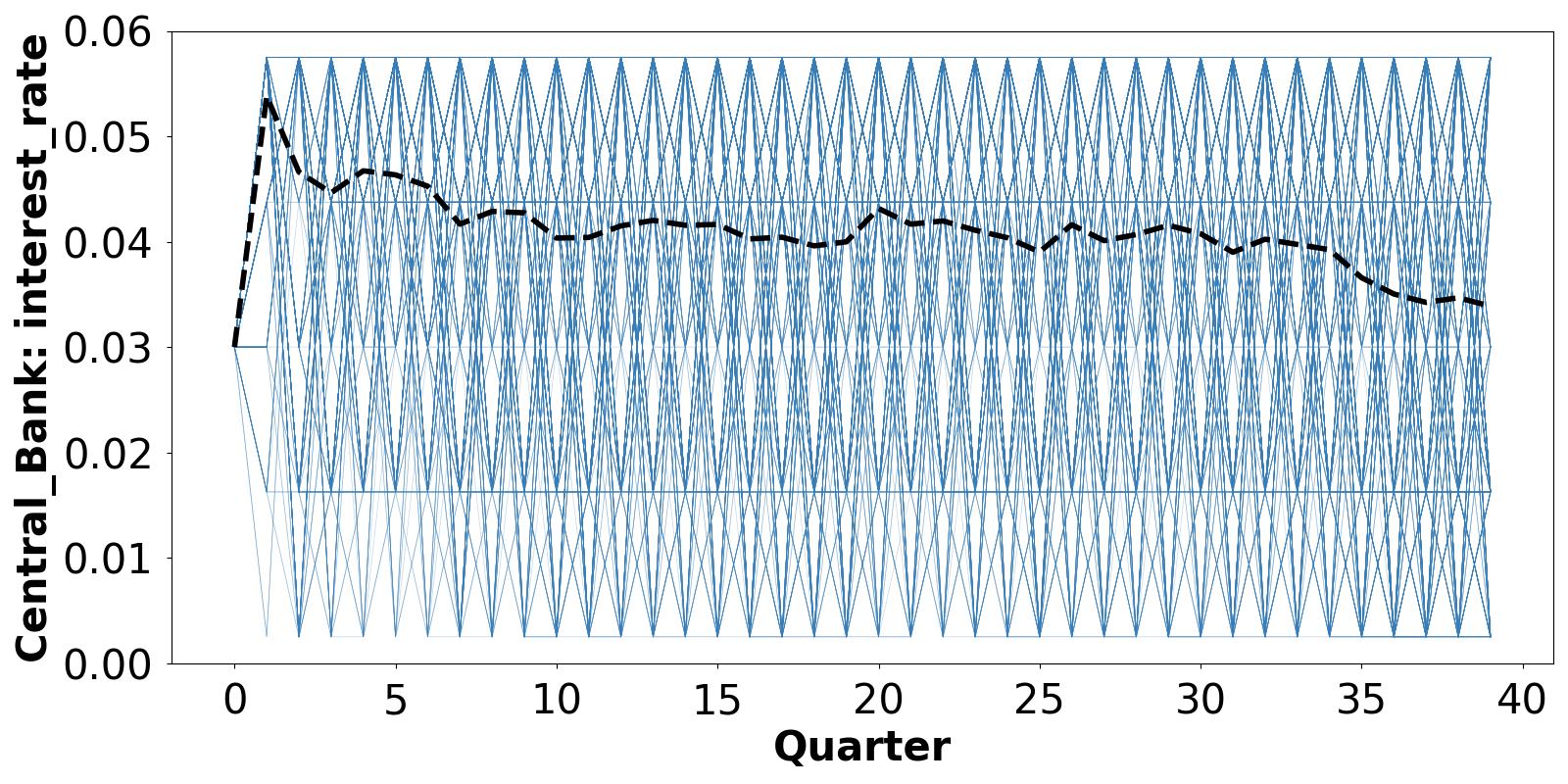}\hspace{0.05\linewidth}
    \includegraphics[width=0.4\linewidth]{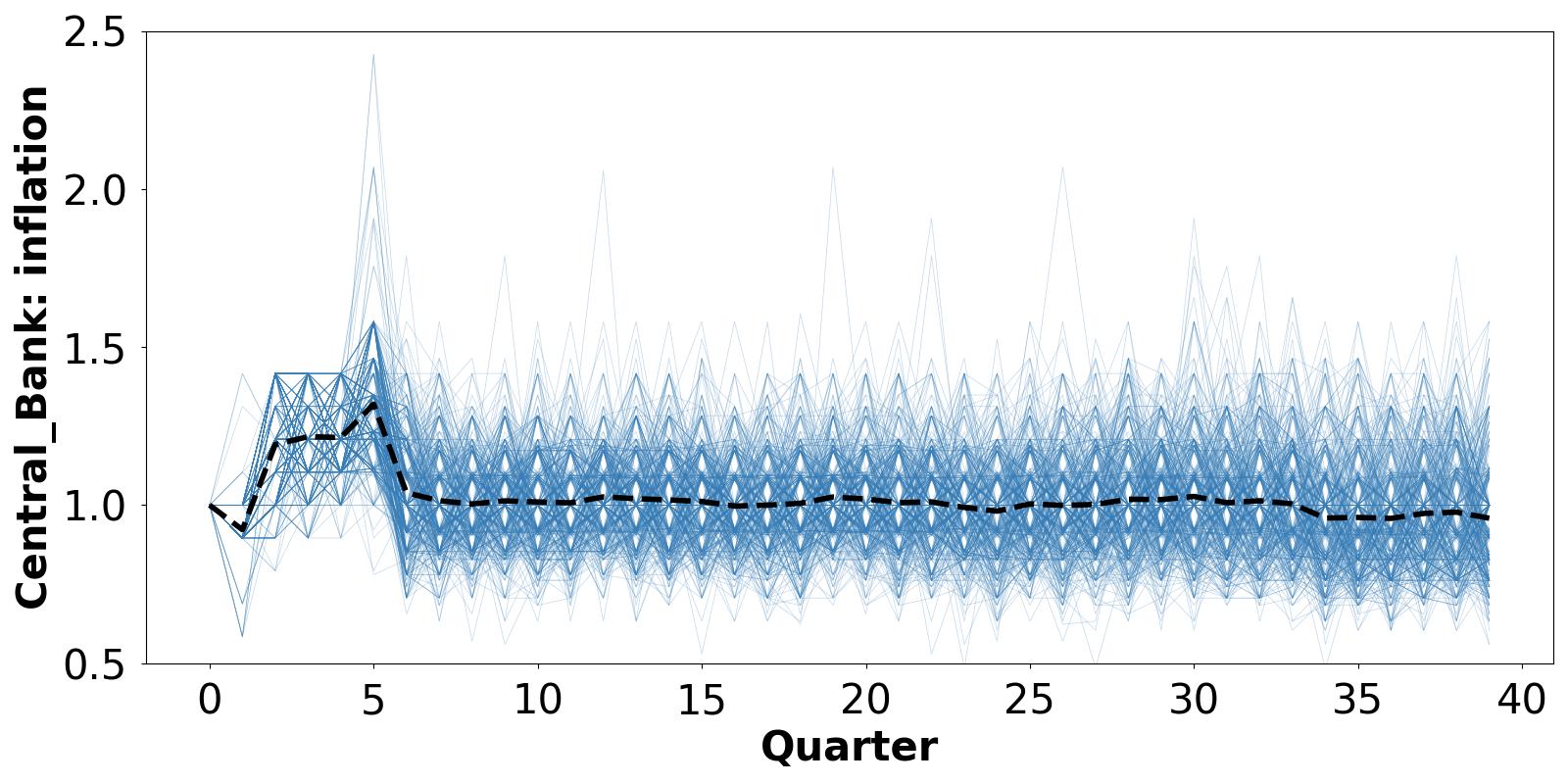}
    \includegraphics[width=0.4\linewidth]{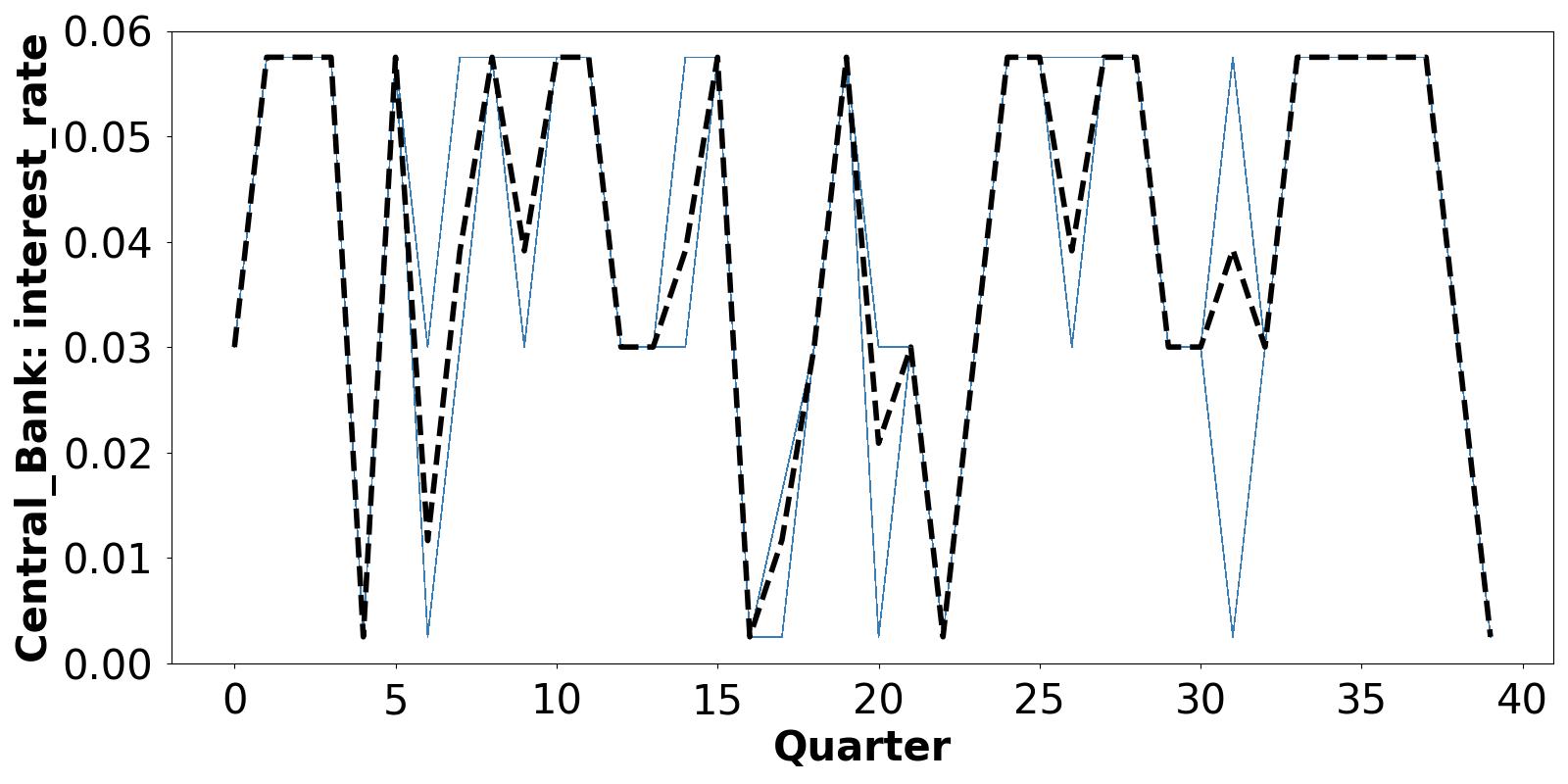}\hspace{0.05\linewidth}
    \includegraphics[width=0.4\linewidth]{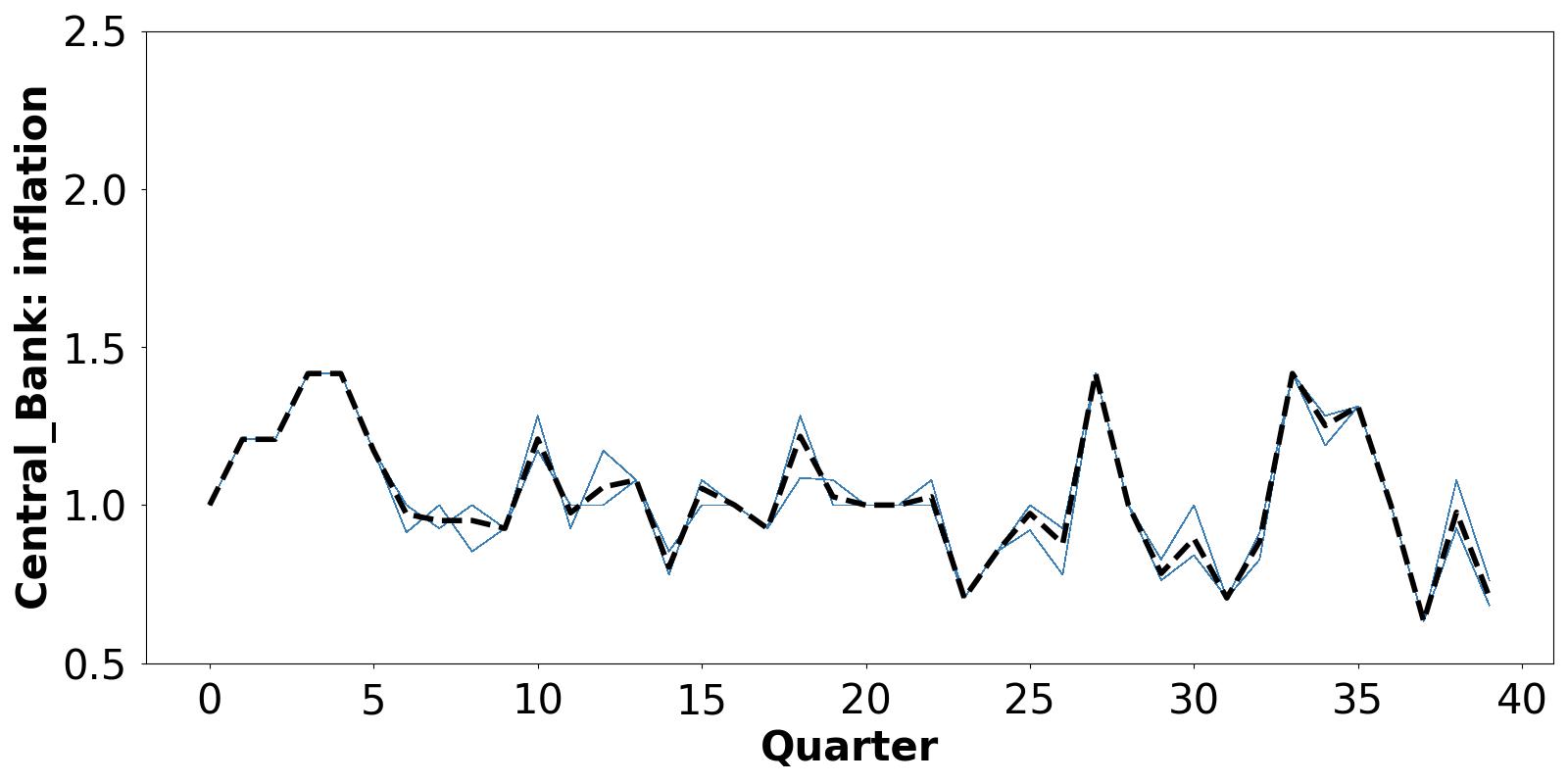}
    \caption{Interest rate (left) and inflation (right) with IMARL (top) and PSRO (bottom) over time. From the bottom row of plots, observe that quarters with high inflation see the setting of high interest rates and vice-versa.}
    \label{fig:app_cb_policy}
\end{figure}

Figure \ref{fig:app_cb_policy} shows the temporal evolution of interest rate set by the central bank (for the next quarter) on the left in response to inflation (in the current quarter) on the right for IMARL (top) and PSRO (bottom). The blue traces show observations per test episode with the black line showing the average across test episodes. From the bottom row of plots, we observe that quarters with high inflation (relative to $\pi^\star=1.02$) see the setting of high interest rates, and those with low inflation see the setting of low interest rates. This is in line with standard monetary policy rules that express the interest rate set by the Central Bank as an increasing function of inflation \cite{taylor2010simple}.

\begin{figure}[tb]
    \centering
    \includegraphics[width=0.6\linewidth]{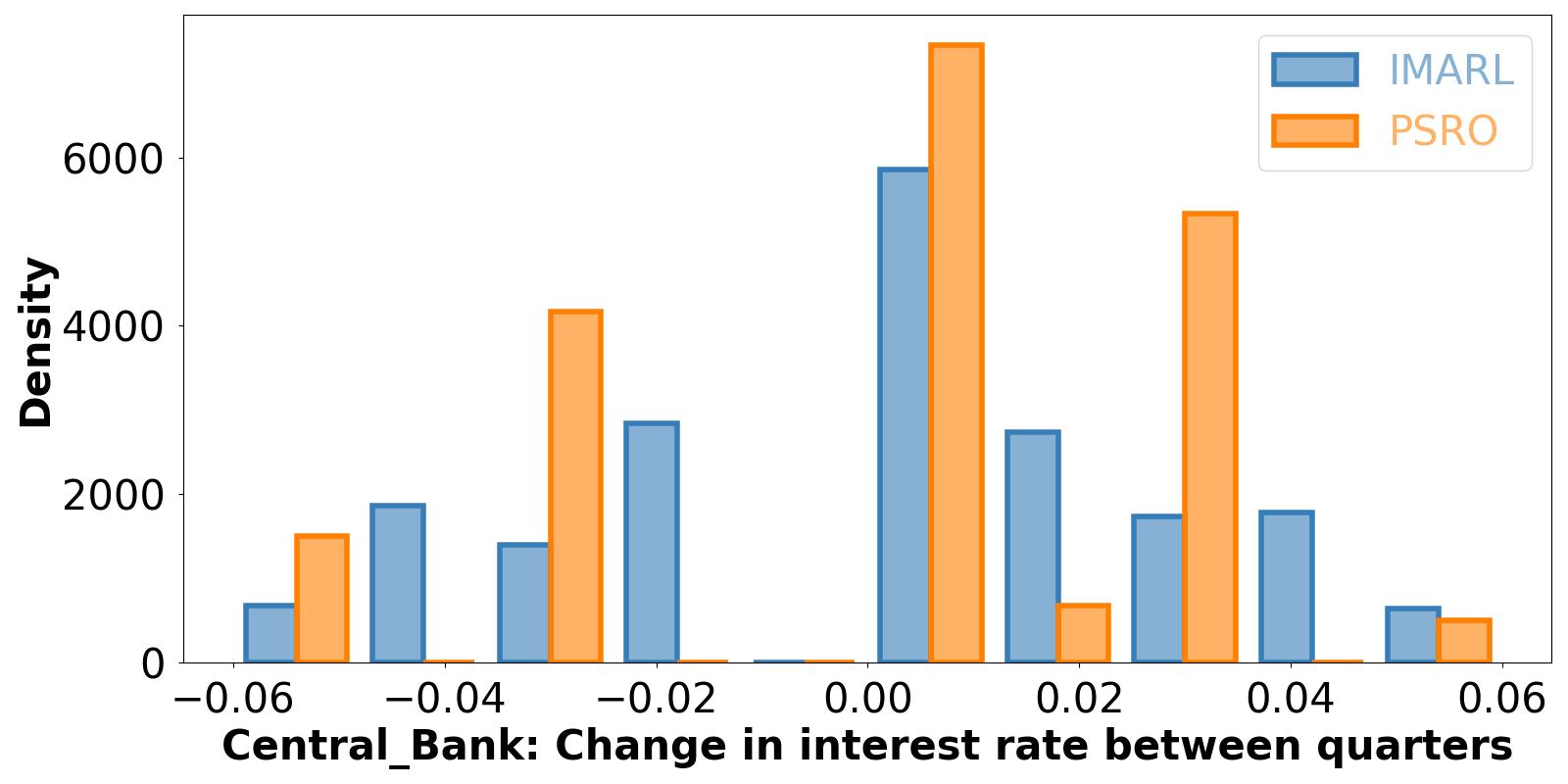}
    \caption{Distribution of change in interest rate between quarters with IMARL and PSRO. With PSRO, rates change less frequently, but by larger amounts when they do.
    }
    \label{fig:app_cb_diff_rates}
\end{figure}

In addition, we observe traces where inflation is very high with IMARL whereas it is more contained with PSRO. Also see that the central bank seems to change rates more frequently with IMARL. To investigate this further, we plot the distribution of change in interest rates from one quarter to the next (across test episodes) in Figure \ref{fig:app_cb_diff_rates}. See that PSRO has higher density than IMARL in the bin for 0 change. However as observed in section \ref{subsec:comp_policies}, when the central bank does alter rates with PSRO, the changes are larger, as it tends to avoid the intermediate rates of 1.625\% and 4.375\%.


\subsection*{Regret metric}
\begin{figure}[h]
    \centering
    \includegraphics[width=0.75\linewidth]{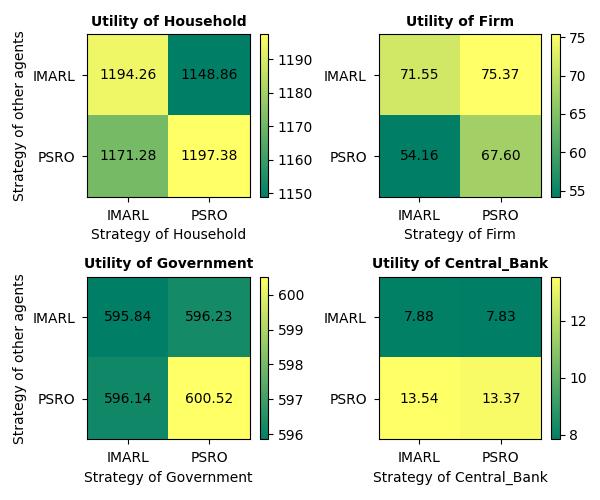}
    \caption{Agent utilities used in computing regret of IMARL and PSRO policies. Regret is given by the difference between off-diagonal and diagonal entries in each row when positive, and zero otherwise.
    }
    \label{fig:regret}
\end{figure}

The regret metric is derived from agent utilities for each joint policy, where utility is the average of an agent's discounted cumulative rewards over 100 test episodes. Figure \ref{fig:regret} shows utilities of each of the four agent types, when they unilaterally deviate from their IMARL/PSRO policy to the other. These utilities are used to compute the regret metrics in Table \ref{tab:regret}. For example, the regret of IMARL for households is calculated as follows: when all other agents adhere to their IMARL policies (top row), households gain no utility by switching from their IMARL policy (left column) to their PSRO policy (right column), resulting in a regret of 0. However, firms and the government experience positive regret from the IMARL policy.

\end{document}